\documentclass[american,prX,longbibliography]{revtex4-1}
\usepackage[T1]{fontenc}
\usepackage[utf8]{inputenc}
\usepackage{booktabs}
\usepackage{amsmath}
\usepackage{amssymb}
\usepackage{graphicx}
\usepackage{color}
\usepackage{float} 
\makeatletter

\providecommand{\tabularnewline}{\\}

\makeatother

\usepackage{babel}
\begin{document}
\global\long\def\ket#1{\left|#1\right\rangle }%

\global\long\def\bra#1{\left\langle #1\right|}%

\global\long\def\ketL#1{\left.\left|#1\right\rangle \right\rangle }%

\global\long\def\braL#1{\left\langle \left\langle #1\right|\right.}%

\global\long\def\braket#1#2{\left\langle #1\left|#2\right.\right\rangle }%

\global\long\def\ketbra#1#2{\left|#1\right\rangle \left\langle #2\right|}%

\global\long\def\braOket#1#2#3{\left\langle #1\left|#2\right|#3\right\rangle }%

\global\long\def\mc#1{\mathcal{#1}}%

\global\long\def\nrm#1{\left\Vert #1\right\Vert }%

\title{Drift-resilient mid-circuit measurement and state preparation error mitigation for dynamic circuits}
\begin{abstract}
Quantum error mitigation (QEM) for dynamic circuits, i.e., those incorporating mid-circuit measurements and feedforward, is important for two key reasons. First, quantum error correction (QEC) circuits are instances of dynamic circuits, and therefore a dynamic circuit-compatible QEM can extend circuit depth and address errors that QEC struggles with. Second, recent studies show that dynamic circuits can significantly outperform purely unitary ones. However, mid-circuit measurement errors remain a major bottleneck. Current solutions rely on readout noise characterization that is vulnerable to temporal noise drifts. To the best of our knowledge, no readout mitigation schemes are resilient
to temporal noise drifts. By introducing parity-based noise amplification in repeated measurements, we derive and experimentally demonstrate a drift-resilient protocol for addressing preparation, mid-circuit, and terminating measurement errors without requiring calibration or characterization. Drift resilience increases the longest possible execution time (in terms of shots) and enables flexibility by combining data from non-consecutive times. For platforms such as trapped ions, where the measurements are highly disruptive, we provide an alternative reset-based mitigation scheme. We demonstrate our methods experimentally on IBMQ and Quantinuum hardware. Combined with the Layered-KIK gate error mitigation protocol, the presented readout mitigation approach enables "End-to-end" mitigation for dynamic circuits, that can improve the outcomes of QEC experiments, and that covers the widest range of errors to the best of our knowledge. Other applications of the presented methods include a faster alternative to gate-set tomography and  diagnostics of defective qubits during the execution of the target algorithm.\end{abstract}
\author{Jader P. Santos}
\author{Raam Uzdin}
\email{raam@mail.huji.ac.il}

\affiliation{\selectlanguage{american}%
	The Hebrew University of Jerusalem, Jerusalem, Israel 9190401}
\maketitle

\section{Introduction}

Dynamic circuits are emerging as a powerful tool for reducing depth
and  achieving results that are beyond the reach of unitary evolution.
Recent examples of dynamic circuits include efficient and shallow
entangled state preparation  \cite{Wallraff2024MCMprepFanOut,baumer2024LongRange,Circa2024MCMpred,Buhrman2024MCMprep,yu2024MCMprep,Cirac2021adaptiveCircPrep},
gate teleportation \cite{baumer2024LongRange}, quantum Fourier transform
\cite{baumer2024scQFT} and more. Dynamic circuits have also been suggested to
 facilitate variational quantum algorithms (VQAs) that are
free from barren plateaus \cite{deshpande2024dynamicNOplateau,xie202mid-circBarren}.

Essentially, dynamic circuits, also known as adaptive circuits, are
circuits in which part of the system is measured, and the outcome determines
the subsequent operations, that is, feedforward. As such, dynamic circuits
encompass all quantum error correction (QEC) codes. Despite their
advantages, dynamic circuits remain severely limited by noise. In \cite{baumer2024LongRange,Quantinuum2023MCMbeyondUnitary,Wallraff2024MCMprepFanOut,baumer2024scQFT},
mid-circuit measurement errors were not mitigated and therefore contributed significantly
 to the error budget \cite{Shirizly2025RBdynamicCirc}. While gate error mitigation and terminating readout error mitigation have been extensively studied, mid-circuit measurement is a relatively new capability that has not yet received much attention in the mitigation community.

Currently, in the absence of large-scale functional error correction
schemes, noise is addressed using quantum error mitigation (QEM)
techniques. QEM is a post-processing framework that utilizes multiple
noisy variants of the original circuit to estimate noiseless expectation
values. QEM can be applied to both gate errors and readout errors. Presently,
nearly every major quantum computer experiment employs one or more
QEM techniques \foreignlanguage{american}{\cite{huggins2021virtual,kim2021scalable,kandala2019error,song2019quantum,arute2020hartree,urbanek2021mitigating,zhang2020error,sagastizabal2019experimental,Quantinuum2025QuantumMagnetismPEA,kim2023evidence,kim2024IBMstabilizeNoisePEC}}.
Although QEM is sometimes viewed as a short-term solution until QEC
becomes operational, it has been proposed that the combination of QEM and QEC
may remain beneficial in the long term \foreignlanguage{american}{\cite{suzuki2022quantum,2021errorQEMforTgatesQEC,lostaglio2021error,Qedma2025ImportanceConcept,wahl2023ZNEforQEC}}.
In particular, in \cite{LayeredKIK}, it has been suggested that QEM protocols applicable to all dynamic circuits allow
for seamless integration with error correction. In this scheme, QEC addresses
strong uncorrelated noise, while QEM tackles correlated noise, leakage, and
other errors that QEC struggles to correct. The compatibility of various
QEM schemes with dynamic circuits was discussed in \cite{LayeredKIK}, where a new dynamic circuit-compatible QEM protocol for gate errors was introduced.

In the dynamic circuit error mitigation scheme studied in \cite{LayeredKIK},
the mid-circuit measurements were assumed to be ideal. Therefore,
mid-circuit measurement errors necessitate an additional mitigation protocol.
Unlike terminating measurement error, mid-circuit error mitigation
must be carried out as quickly as possible to avoid quantum state
degradation during classical processing. A protocol based on the
 characterization of measurement errors and probabilistic error cancellation was presented
in \cite{hashim2025PECreadout,koh2024readout,Temme2024dynamicPEC}. However, all characterization-based
methods are intrinsically vulnerable to temporal variations in noise parameters
during the course of the experiment, i.e., noise drifts. This sensitivity
becomes increasingly problematic at higher target accuracies that
require longer and more frequent characterization periods. Moreover, QEM incurs substantial  runtime overhead, which increases the likelihood of experiencing non-negligible noise drift effects. Such
drift-sensitivity is also an issue in terminating measurement readout error mitigation.
To the best of our knowledge, there is no drift-resilient readout error mitigation for terminating or  mid-circuit measurements. In this work, we introduce such schemes. 

Importantly, in mid-circuit measurements, there is a distinction between
the readout noise that manifests in the measurement outcomes and
the noise that degrades the state of the qubits during the measurement,
 e.g., the decay of the qubits to their ground state. Currently, no mitigation
technique simultaneously addresses both types of errors. This is another
gap addressed by our framework.

Sensitivity to noise drift also poses a challenge in gate mitigation. While
some methods, such as PEC \cite{temme2017ZNEPECorig} and PEA \cite{kim2023evidence}, face the same drift sensitivity problem since they require an updated
noise model, other methods, such as purification \cite{Koczor2021},
KIK \cite{npjqiKIK,LayeredKIK}, and pulse-stretching zero noise extrapolation
(ZNE) \cite{temme2017ZNEPECorig}, are drift-resilient. In the KIK
approach, the noise is amplified by amplification factors of $2j+1$, 
where $j=0,1,2,\ldots$  By combining results obtained under different amplification
levels, one can reconstruct the inverse noise channel in post-processing.
Because this method does not require noise characterization, circuit execution order 
can be chosen to ensure complete robustness against time-dependent
noise drift.

Motivated by the noise-agnostic amplification approach employed in KIK
and ZNE to address gate errors, our objective is to mitigate measurement
errors by amplifying them in positive odd powers. In gate mitigation,
error amplification can be achieved by repeating a gate and its inverse;
however, this strategy fails for measurement noise. As mentioned above,
measurement error has two principal components: (i) pure readout error,
which occurs even when the population in the measurement basis remains
unchanged, and (ii) population degradation, where the qubit state
changes during measurement, e.g., due to spontaneous decay. Although the
second type of noise is amplified by sequentially repeating the measurement protocol, the first type is not. If the qubit populations in the
measured basis remain unchanged throughout the measurement, repeating
the measurement merely increases the number of samples, each affected
by the same readout noise as the first measurement. 

In this work, we demonstrate that processing the joint outcomes
of a sequence of measurements enables controlled
amplification of measurement noise. In the absence of population degradation
(e.g., decay to the ground state), we analytically show that the classical
parity operation (XOR) yields the desired noise amplification. 
When qubit population errors are pronounced, we propose two alternative
protocols. The first is based on inserting dummy measurements, while
the second assigns state-dependent weights to different outcomes.
For terminating measurements, both protocols offer similar mitigation
performance, but differ in implementation complexity. In the
case of mid-circuit measurements, the latter protocol, termed "weighted
parity", provides stronger mitigation of qubit degradation noise, including spectator qubits, due to its shorter runtime. Nevertheless,
the dummy measurement is easier to understand and code. Finally, for
platforms such as trapped ions, where the measurement substantially
alters the qubit state but a high-quality reset is available, we introduce
an alternative noise amplification based on reset operations and feedforward.

\subsection{Measurement errors}

Consider a single-qubit scenario where the state of the qubit is $q\ketbra 11+(1-q)\ketbra 00$
which we write as $\{q,1-q\}$ or just $q$ . In each shot, the error-free
outcome is  \textquotesingle $1$\textquotesingle$=\{1,0\}$ or \textquotesingle $0$\textquotesingle$=\{0,1\}$.
In addition, there is the readout error that we model using a stochastic
matrix $M$ that operates on $\{q,1-q\}$ so that the observed distribution
is $\{p,1-p\}=M\{q,1-q\}$. As before, in each shot, the readout outcome
is either \textquotesingle $0$\textquotesingle \ or \textquotesingle $1$\textquotesingle, but the probabilities of observing these results
are influenced by the readout noise. The modeling of readout
noise using a stochastic matrix $M$ may not be entirely accurate
in some platforms \cite{maciejewski2020QuantumReaedoutMitigation}.
Nonetheless, techniques such as measurement twirling and forced dephasing
improve the accuracy of this description \cite{Temme2024dynamicPEC}.

Standard readout mitigation approaches \cite{M3,van2022modelFreeREM,hashim2025PECreadout,koh2024readout,maciejewski2020QuantumReaedoutMitigation,funcke2022measurementPEC,tannu2022hammer,stein2023Qbeep,smith2021MeasTwirl,kim202MLreadout,Bauer2021BiasFlipMeas,wang2023truncatedVNSeriesReadout,PRX.7.021050BenjaminGenRef,readoutSeries,gunther202readoutHardwareRepetition,shirizly2025feedforwardMeasBias,baron2024qubit,chandramouliBaron2025statistical}
which do not involve hardware overhead (unlike \cite{gunther202readoutHardwareRepetition})
mitigate the probability distribution, but do not mitigate individual
shots. These methods typically utilize matrix inversion approach or alternative techniques to solve $p=Mq$. In the first approach, $M^{-1}$ is applied
to the final noisy readout distribution in post-processing. Given
that readout errors might be correlated, the calculation of $M^{-1}$
is nontrivial for a large number of qubits. This classical overhead
can be reduced using sparsity arguments \cite{M3} and twirling \cite{wallman2016noise,van2022modelFreeREM,smith2021MeasTwirl,Wallman2023MeasurementTwirling,hashim2025PECreadout,koh2024readout}
that simplify the noise structure. In the other approach, the real
distribution $q$ is obtained by iteratively solving the equation
$p=Mq$. In both cases, $M$, or a proxy of it, must be learned, and the quality of the mitigation depends on the accuracy of the characterization
of $M$, as well as on the temporal drift of the matrix elements of
$M$. Depending on the platform and the duration of the experiment,
$M$ may change during the execution of the experiment. This becomes
particularly important in the context of gate error mitigation, which
often incurs significant sampling overhead and can increase the total
runtime by an order of magnitude or more. As a result, noise drifts
are more likely to occur in error-mitigated experiments compared to unmitigated experiments.

In dynamic circuits, the readout in each shot determines which circuit
will be executed next. Thus, in each shot, a string of outcomes in the measurement
basis, such as $s=010100 $, must be generated . When the
measurements are twirled (see discussion in Sec. \ref{subsec: Extension-to-arbitrary}),
the matrix inverse can be expressed using operations that turn one
string into another. This approach has been proposed for mid-circuit
measurement error mitigation and experimentally demonstrated in \cite{hashim2025PECreadout,koh2024readout}. See also \cite{Temme2024dynamicPEC} where mid-circuit measurement-induced crosstalk is considered.
However, in this approach, a characterization of $M$ is needed. While
the twirling enables faster characterization, it still does not provide
drift-resilience. The characterization must be repeated on the time
scale at which the noise drifts. Moreover, each characterization must
be significantly faster than the interval between adjacent characterizations
(low duty cycle). This issue becomes more critical as the target accuracy
increases and the system size grows. In addition, this approach requires
disentangling readout errors from preparation errors in platforms
where preparation errors are significant (see experiment in Sec.
VII and Appendix III).

In this work, we construct a drift-resilient, characterization-free
mid-circuit measurement mitigation scheme. That is, this method eliminates the need for repeated calibration 
or characterization of the detectors. 
As pointed out in \cite{npjqiKIK,LayeredKIK}, the key to drift-resilience
is to divide the runtime into small time bins in which the noise hardly
changes and apply  a full noise-agnostic (characterization-free)
error mitigation protocol in each bin. To the best of our knowledge, the only
way to achieve noise-agnostic mitigation is by using agnostic noise amplification. The main drawback
of our approach is its longer circuit execution time, compared to \cite{hashim2025PECreadout,koh2024readout,Temme2024dynamicPEC}. Yet, our approach does not require noise characterizations, so when the characterization is slow, our method can be faster overall.

Moreover, we point out that when using PEC-like approaches as in \cite{hashim2025PECreadout,koh2024readout,Temme2024dynamicPEC},
it is possible to fully correct either the readout error or the error associated with the degradation of the qubit state
, but not both. Our approach addresses both issues simultaneously.
In addition, in platforms where preparation errors arise from conditional
reset, our approach mitigates these errors without
incurring additional overhead beyond that required for readout error
mitigation.

The paper is organized as follows. Section \ref{sec: Parity-based-measurement} begins
with a motivating single-qubit example that excludes decay effects
and shows that by evaluating the parity of consecutive measurements,
it is possible to reliably amplify readout errors. Next, we generalize
this analytically to any number of qubits and any form of (classical)
correlation between their readout errors. In Sec. \ref{sec: Accounting-for-the}, 
we study the effect of decay and devise two methods to eliminate its
first-order impact. One method uses dummy measurements, and the other
employs weighted parity. A variant suitable for platforms with high-quality 
reset is presented in Sec. IV. In Sec. V, we extend
our approach to address state preparation errors. In Sec. VI, we describe
the integration of our method with other readout and gate mitigation
techniques, and in Sec. VII we present experimental results. Finally,
in Sec. VIII, we conclude and discuss two side applications of the
current approach. To avoid cumbersome expressions and derivations,
different derivations throughout the paper require alternative notations
and formalisms.

\section{Parity based measurement error amplification}\label{sec: Parity-based-measurement}

\subsection{Mitigation by noise amplification}

Assume for a moment that it is possible to generate measurements whose assignment matrices are
of the form $M^{3},M^{5},...$ and so on. Next, we perform measurements
using $M^{2j+1}$ and obtain the distributions $p^{(j)}=M^{2j+1}q$,
where $q$ is the probability vector of the population in the multi-qubit computational basis.
By combining $p^{(j)}$ with specific weights, it is possible to exponentially
suppress the noise. Using Taylor coefficients \cite{npjqiKIK}, also known as Richardson coefficients,  
\begin{equation}
a_{j}^{(m)}=\frac{(-1)^{j}(2m+1)!!}{2^{m}(2j+1)j!(m-j)!},
\end{equation}
and writing $M$ as $M=e^{\epsilon A}$, where the scalar $\epsilon$
captures the strength of the readout error, we obtain the mitigated
distribution
\begin{equation}
p_{mit}^{(m)}\doteq\sum_{k=0}^{m}a_{k}^{(m)}p^{(k)}=\sum_{k=0}^{m}a_{k}^{(m)}M^{2k+1}q=q+O(\epsilon^{m+1}).\label{eq: p(m) mit}
\end{equation}
It is more convenient to express this result in terms of the mitigated
assignment matrix

\begin{equation}
M_{mit}^{(m)}\doteq\sum_{k=0}^{m}a_{k}^{(m)}M^{2k+1}=I+O(\epsilon^{m+1}).\label{eq: M(m)_mit}
\end{equation}
The proof follows from Taylor expanding to order $m+1$
around $\epsilon=0$, and using the properties $\sum_{j=0}^{m} a_{j}^{(m)}[(2j+1)\epsilon]^{l}=0$
for $l=1,2,\ldots,m$ and $\sum_{j=0}^m a_{j}^{(m)}=1$ \cite{LayeredKIK}.
As a simple illustrative example, consider a symmetric single-qubit
assignment matrix
\begin{equation}
M=\left(\begin{array}{cc}
1-\epsilon & \epsilon\\
\epsilon & 1-\epsilon
\end{array}\right)=I+\epsilon(\sigma_{x}-I).\label{eq: symM1q}
\end{equation}
Then,
\begin{align}
M^{3} & =I+3\epsilon(\sigma_{x}-I)+3\epsilon^{2}(\sigma_{x}-I)^{2}+\epsilon^{3}(\sigma_{x}-I)^{3},\label{eq: M^3}
\end{align}
and the mitigated measurement operators are 
\begin{align}
M_{mit}^{(1)} & =\frac{3}{2}M-\frac{1}{2}M^{3}=I+(3\epsilon^{2}-2\epsilon^{3})(\sigma_{x}-I)=I+O(\epsilon^{2}),\\
M_{mit}^{(2)} & =\frac{15}{8}M-\frac{5}{4}M^{3}+\frac{3}{8}M^{5}=I+(10\epsilon^{3}-15\epsilon^{4}+6\epsilon^{5})(\sigma_{x}-I),\nonumber \\
 & =I+O(\epsilon^{3}).
\end{align}
So far, we have only used known mathematical properties of the Taylor
coefficients \cite{LayeredKIK}. In the next section, we
introduce an operational protocol for generating measurement outcomes
whose noise is amplified by a factor of $2j+1$. The drift-resilience
of this approach does not depend on the details of the amplification
as long as the amplification is agnostic to the details of the noise
and does not involve noise characterization.  For quantification of drift-resilience, see the last part of
Appendix I in \cite{LayeredKIK}.

Series-based readout mitigation has also been suggested in \cite{readoutSeries,wang2023truncatedVNSeriesReadout}.
However, the method in \cite{readoutSeries} requires characterization
and is therefore not drift-resilient, and in \cite{wang2023truncatedVNSeriesReadout}
no operational protocol was suggested to implement the noise amplification.

\subsection{Introducing the parity-based noise amplification}

Next, we consider a sequence of three measurements of the same qubit
in the same shot. For now, we neglect decay effects, ensuring that each measurement
experiences the same qubit state. For simplicity of presentation,
we first consider the symmetric assignment matrix in eq. (\ref{eq: symM1q}),
and assume the qubit state is \textquotesingle $1$\textquotesingle. Table \ref{tab: decay free} shows the
probability of observing each possible sequence of measurement outcomes.
\begin{table}
\begin{tabular}{ccccc}
\toprule 
$m_{1}$ & $m_{2}$ & $m_{3}$ & Probability & Parity\tabularnewline
\midrule
1 & 1 & 1 & $(1-\epsilon)^{3}$ & 1\tabularnewline
1 & 0 & 0 & $\epsilon^{2}(1-\epsilon)$ & 1\tabularnewline
0 & 1 & 0 & $\epsilon^{2}(1-\epsilon)$ & 1\tabularnewline
0 & 0 & 1 & $\epsilon^{2}(1-\epsilon)$ & 1\tabularnewline
0 & 0 & 0 & $\epsilon^{3}$ & 0\tabularnewline
0 & 1 & 1 & $\epsilon(1-\epsilon)^{2}$ & 0\tabularnewline
1 & 0 & 1 & $\epsilon(1-\epsilon)^{2}$ & 0\tabularnewline
1 & 1 & 0 & $\epsilon(1-\epsilon)^{2}$ & 0\tabularnewline
\bottomrule
\end{tabular}

\caption{The probability of observing different outcomes in a sequence of
three measurements $\{m_{1},m_{2},m_{3}\}$ where the qubit state is
\textquotesingle $1$\textquotesingle \  and the readout error is $\epsilon$. }\label{tab: decay free}
\end{table}
The measurements are assumed to be statistically independent. By summing
over all mutually exclusive events corresponding to parity one $\{111,100,010,001\}$
we obtain the probability of measuring parity \textquotesingle $1$\textquotesingle \ 
\begin{equation}
\mathbb{P}(\text{Parity}=1|q=1)=(1-\epsilon)^{3}+3\epsilon^{2}(1-\epsilon),
\end{equation}
which is exactly the probability of observing \textquotesingle $1$\textquotesingle \ in $p^{(3)}$: 
\begin{equation}
p^{(3)}=M^{3}\left(\begin{array}{c}
1\\
0
\end{array}\right)=\left(\begin{array}{c}
(1-\epsilon)^{3}+3\epsilon^{2}(1-\epsilon)\\
3\epsilon-6\epsilon^{2}+4\epsilon^{3}
\end{array}\right).
\end{equation}
The same occurs when the qubit state is \textquotesingle $0$\textquotesingle. Parity classifies as
\textquotesingle $1$\textquotesingle (\textquotesingle $0$\textquotesingle) states such as $001$ ($101$) that are more likely to be
associated with the incorrect qubit state \textquotesingle $0$\textquotesingle (\textquotesingle $1$\textquotesingle). This demonstrates
how the parity of a sequence amplifies the readout error in a controllable manner with respect
to a single-measurement error. As we quantify next, parity of longer sequences
involve more misclassifications, further amplifying the readout
error in a systematic manner. 
 At this point, the reader may wonder why we amplify the noise instead of using  majority vote to reduce it. Later, we study the majority vote protocol and show both theoretically and experimentally that it leads to poor performance. However, this becomes apparent only after taking into account the effect of qubit decay during the measurement as explained in Sec. \ref{sec: Accounting-for-the}.

More generally, we claim that the noise-amplified distribution $M^{2j+1}q$
can be replaced by the distribution of outcomes generated by the parity of $2j+1$ consecutive
measurements of $q$, i.e.,  $par(m_{1},m_{2},\ldots,m_{2j+1})=\oplus_{i=1}^{2j+1}m_{i}$ , where $\oplus$ denotes the classical XOR operation. As shown next, when applying measurement twirling, this parity principle
holds for any number of qubits and any correlated measurement errors
(within the stochastic description of readout errors).  Finally, we point that sequential measurements have also been used to characterize leakage errors \cite{Dai2025repeatedMeasLeakage}.
\subsection{Extension to arbitrary number of qubits and correlated measurement
error}\label{subsec: Extension-to-arbitrary}

\begin{figure}
\includegraphics[width=\textwidth]{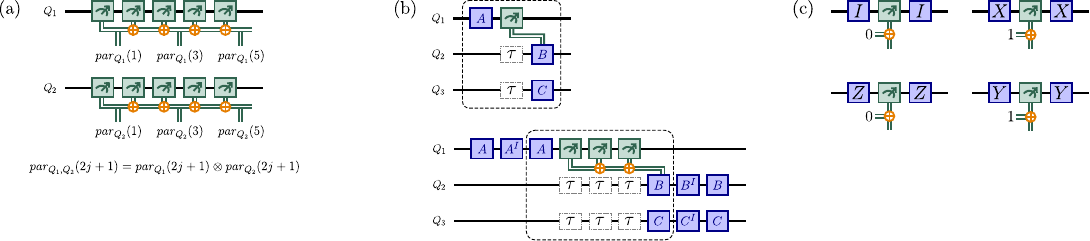}
\caption{(a) A schematic description of the parity-based readout noise amplification.
In each shot, each qubit ($Q_{1}$ and $Q_{2}$ in this example), the
parity of $2j+1$ consecutive measurements is used for amplifying the
readout assignment matrix $M$ by a factor of $2j+1$, i.e., $M\to M^{2j+1}$.
The parity is evaluated for each qubit separately, and the multi-qubit
parity outcome is the tensor product of the individual parity outcomes.
When applied to terminating measurements, the measurements used for
amplification level five, for example, can be used also for amplification
level three. That is, separate circuits are not required for different
amplification factors. (b) In mid-circuit measurements with feedforward operation 
(dashed box), different amplification factors produce different feedforward
outputs, and therefore, each amplification factor requires a separate
circuit. The delay, both for the measured qubit and the spectator
qubits during the sequence is also scaled by a factor of $2j+1$.
As a result, qubit degradation due to this delay is automatically
mitigated for spectator qubits ($Q_{3}$) as well. To the
best of our knowledge, this feature is unique to our approach. Finally,
the operations outside the box show how the proposed readout mitigation
scheme fits well with the Layered KIK gate mitigation scheme. The same
amplification factor is applied to both the gates and the measurements
("uniform amplification"). (c) Illustration of the four single-qubit
measurement twirlings. If the input twirling Pauli is X or Y (Z or
I) a bit flip is (not) applied in post-processing as denoted by the
$\oplus1$ ($\oplus0$) operation.}\label{fig: parity no decay}
\end{figure}
The key element in our protocol is illustrated for two qubits and
amplification level of three in Fig. \ref{fig: parity no decay}(a).
Importantly, this is not the complete form of protocol as it does
not properly account for noise in the qubit space (e.g., amplitude
damping). In each shot, the parity of the first $2j+1$ measurements
is recorded for each qubit. The multi-qubit outcome is the tensor
product of the parity outcome of each qubit. The normalized frequencies
(counts) of the parity outcomes yield $p^{(2j+1)}$. 

For terminating measurements and without gate error amplification,
the measurements used, for example, for amplification factor five,
can also be used for amplification factors three and one [see Fig. 1(a)] . The resulting
statistical dependence does not affect the average. In first-order
mitigation, one can show that the covariance actually helps as it
reduces the variance of the mitigated result. The point will be 
discussed further in Sec. \ref{sec: Accounting-for-the}. However, when the
measurement outcome is used for feedforward, as shown in the dashed
box in Fig. \ref{fig: parity no decay}b, and (or) when agnostic amplification
gate error mitigation (operation outside the box) is integrated with
the readout mitigation, each amplification factor has to be carried
out in a separate circuit. 

In Appendix I, we consider the case of an arbitrary stochastic assignment matrix $M$ 
and show that parity-based noise amplification holds even when $M$
has no special symmetries. Yet, measurement twirling \cite{wallman2016noise,smith2021MeasTwirl,Wallman2023MeasurementTwirling,hashim2025PECreadout,koh2024readout}
leads to a special structure that is easy to learn if needed, and
it also reduces coherent non-Markovian effects such as crosstalk with
spectator qubits. Hence, in the main text, we focus  on the case where $M$
is twirled, and refer the reader to Appendix I for the more general
case. In what follows, we prove that when $M$ is twirled, the distribution
of samples generated from $M^{(2j+1)}q$  is identical to the distribution
generated by evaluating the parity of $2j+1$ consecutive measurements
for each shot. 

A general matrix can be written in the multi-qubit Pauli basis
\begin{equation}
M=\sum_{j_{i}\in\{0,1,2,3\}}\alpha_{j_{1},j_{2},j_{3}\ldots}\sigma_{j_{1}}\otimes\sigma_{j_{2}}\otimes\sigma_{j_{3}}\otimes \ldots
\end{equation}
Measurement twirling for mid-circuit measurement is shown in Fig. \ref{fig: parity no decay}c.
A random Pauli gate is physically executed before and after the measurement,
and another Pauli gate is executed in post-processing to correct the
change made by the first Pauli. This procedure is performed for each
measurement in the sequence separately, and the final twirled assignment matrix $M_{twirled}$ is obtained by averaging over different random Pauli
choices. 

After applying measurement twirling \cite{wallman2016noise,smith2021MeasTwirl,Wallman2023MeasurementTwirling,hashim2025PECreadout,koh2024readout}
only elements that contain solely $\sigma_{0}$ or $\sigma_{x}$,
e.g., $\sigma_{x}\otimes\sigma_{0}\otimes\sigma_{x}\otimes\sigma_{x}\otimes\sigma_{0}$,
remain. Thus, $M_{twirled}$ of $n$ qubits can be interpreted as noise composed
of bit flips only. 
\begin{equation}
M_{twirled}=\sum_{j_{i}\in\{0,1\}}\alpha_{j_{1},j_{2},j_{3},\ldots}\sigma_{j_{1}}\otimes\sigma_{j_{2}}\otimes\sigma_{j_{3}}\otimes\ldots
\end{equation}
At this point, it is useful to describe the system in terms of outcome
strings rather than quantum states. The string $s=(1,1,0,...)$ implies
that qubits number 1 and 2 are in the state \textquotesingle $1$\textquotesingle, the third qubit is
in the state \textquotesingle $0$\textquotesingle \ and so on. To apply a bit flip, we XOR the string with
1; to leave it unchanged, we XOR it with 0. For example, the
application $\sigma_{x}\otimes\sigma_{0}\otimes\sigma_{x}\otimes\sigma_{0}\otimes\sigma_{0}$
on a state q, is obtained by XOR-ing $s_{q}$ with $(1,0,1,0,0)$. We conclude that
in the string formalism, the measurement error can be written as
\begin{equation}
p=M_{twirled}q\iff\{s_{p}\}=M_{s}\oplus s_{q},
\end{equation}
where $M_{s}\oplus =\sum_{j_{i}\in\{0,1\}}\alpha_{j_{1},j_{2},...,j_{n}}\{j_{1},j_{2},...,j_{n}\}\oplus$
and $s_{q}$ is the string that describes the state of the $n$ measured qubit.
Without loss of generality, we assume that $q$ is one of the measurement
outcomes and not a mixture of them. By $\sum\oplus$ we mean the following,
if we have $[0.3(1,0,1)\oplus+0.1(0,1,0)\oplus](1,0,0)=\{\{0.3,(0,0,1)\},\{0.1,(1,1,0)\}\}$.
Thus, $\{s_{p}\}$ is a set of strings that occur with probabilities
determined by the values of $\alpha$. Next, we express the error-amplified readout state $M_{twirled}^{2j+1}q$ in terms of strings

\begin{equation}
\{s_{M^{2j+1}}\}=[\sum_{k_{i}\in\{0,1\}}\alpha_{k_{1},k_{2},\ldots,k_{n}}\{k_{1},k_{2},\ldots,k_{n}\}]^{\oplus(2j+1)}\oplus s_{q},
\end{equation}
and compare this expression to the parity of $2j+1$ consecutive measurements operating on the same $s_{q}$. The parity operation can be written as
\begin{equation}
\{s_{par({2j+1)}}\}=[\sum_{k_{i}\in\{0,1\}}\alpha_{k_{1},k_{2},\ldots,k_{n}}\{k_{1},k_{2},\ldots,k_{n}\}\oplus s_{q}]^{\oplus(2j+1)}.\label{eq:  spar}
\end{equation}
Since for now we assume that  the qubit does not decay during the measurement,
 the same $s_{q}$ appears in all the $2j+1$ measurements.
Since the XOR operation is commutative, we can rearrange Eq. (\ref{eq:  spar})
into the form
\begin{equation}
\{s_{par({2j+1)}}\}=[\sum_{k_{i}\in\{0,1\}}\alpha_{k_{1},k_{2},\ldots,k_{n}}\{k_{1},k_{2},\ldots,k_{n}\}]^{\oplus(2j+1)}s_{q}^{\oplus(2j+1)}.
\end{equation}
However, since we are interested in odd powers, $s_{q}^{\oplus(2j+1)}=s_{q}$
and we finally get 
\begin{align}
\{s_{par({2j+1)}}\}=[\sum_{k_{i}\in\{0,1\}}\alpha_{k_{1},k_{2},\ldots,k_{n}}\{k_{1},k_{2},\ldots,k_{n}\}]^{\oplus(2j+1)}s_{q}
=\{s_{M^{2j+1}}\},
\end{align}
which completes the proof for an arbitrary number of qubits and any correlation in the readout error of different qubits. To clarify the operational protocol, consider three consecutive four-qubit readout outcomes for a specific shot $i$, $m_{1}^{(i)}=1001,m_{2}^{(i)}=1011,m_{3}^{(i)}=1110$.
 The parity is evaluated for each qubit separately $m_{1}\oplus m_{2}\oplus m_{3}=(1\oplus1\oplus1,0\oplus0\oplus1,0\oplus1\oplus1,1\oplus1\oplus0)=(1,1,0,0)$.
In table form, we have
\begin{equation}
\begin{array}{c}
qubit\#\downarrow\end{array}\begin{array}{cccc}
m_{1}^{(i)} & m_{2}^{(i)} & m_{3}^{(i)} & par(3)\\
1 & 1 & 1 & 1\\
0 & 0 & 1 & 1\\
0 & 1 & 1 & 0\\
1 & 1 & 0 & 0
\end{array}
\end{equation}
The measurement sequence of the $n$-th qubit corresponds to the $n$-th
row. The raw un-amplified shot-averaged result and the three-fold
amplified result are 
\begin{align}
p^{(1)}(m_{1}^{(i)}) & =\#[m_{1}^{(i)}]/N_{shots},\\
p^{(3)} & =\#[m_{1}^{(i)}\oplus m_{2}^{(i)}\oplus m_{3}^{(i)}]/N_{shots}.
\end{align}
where the count $\#[m]$ denotes the number of shots in which the
string (state) $m$ was observed, and $N_{shots}$ is the total number
of shots.

At first, it may appear that the multiple measurements will give the
environment more time to degrade the qubit state (both the measured
and spectator qubits). Although this is correct for each amplification
level separately, such noise is automatically mitigated in post-processing.
As illustrated by the  $\tau $ delay boxes in Fig. \ref{fig: parity no decay}b, the delay also scales like $2j+1$. As a result, any Markovian noise
associated with this delay will be mitigated. For this to apply to spectator qubits
as well, it is important to apply the same delay as the one measured qubits
experience due to the measurements. This is illustrated in \ref{fig: parity no decay}b
by the added delays in the spectator qubit $Q_{3}$. That being said,
because of the decay noise, the consecutive measurement outcomes can no
longer be considered as independent and identically distributed (i.i.d.)
random variables. In the next section, we modify the basic parity-amplification
protocol to address this problem.

\section{Accounting for the qubit decay}\label{sec: Accounting-for-the}

\subsection{The impact of decay on the basic parity protocol}\label{subsec:The-impact-of decay}

In many systems, the more energetic qubit state is meta-stable and
decays after some time to the ground state. We consider the case where
the decay probability $\gamma$ during a single measurement is non-negligible
and degrades the performance of the basic parity-mitigation protocol
discussed above. Yet, as explained later, we also assume that it is
not too strong, so that $O(\gamma^{2})$ corrections must included. This regime is very common in superconducting qubits
for the following reason. In principle, the separation between \textquotesingle $0$\textquotesingle\ 
and \textquotesingle $1$\textquotesingle \ in the IQ plane of the readout resonator increases when the
measurement integration time is longer. However, if the qubit decays
during the integration, the separation is degraded, and the readout
error rate increases. Hence, the integration time is limited by the
decay time. This implies that the decay effects in this scenario are
1) not negligible and 2) not too strong.

To model the decay effect, we assume that before each measurement
there is a certain probability $\gamma$ for the qubit to decay. Placing
the decay after the measurement leads to slightly different results, which we find to be inconsistent with the experiment. Moreover, it
makes the first measurement different from the others. Since our method
for addressing the decay eliminates $O(\gamma)$, our derivation neglects
terms of order $O(\gamma^{2},\gamma\epsilon)$. This implies that
in multi-qubit scenarios, we neglect events in which several qubits
decay simultaneously in a given shot. For a given decay rate, the
probability of simultaneous decay increases with the number of qubits.
Therefore, accounting for order $\gamma^{2}$ is important and warrants
further study. 

We start with the qubit in state \textquotesingle $1$\textquotesingle\ \ and carry out a sequence of three
measurements. The physical state of the qubit is $111$ if no decay occurs,
$110$ if the decay occurs just before the third measurement, $100$ if
it occurs before the second, and $000$ if it occurs before the first.
Since, for now, we are considering only decay noise, a transition back
to the \textquotesingle $1$\textquotesingle\  state is not possible. Moreover, the probability of measuring,
for example, $111$, after the qubit decays to $110$ is neglected because it
is of order $O(\gamma\epsilon)$. For a single measurement, we have
\begin{equation}
\mathbb{P}(par(1)=1)=(1-\gamma)(1-\epsilon)+\gamma\epsilon=(1-\epsilon)-\gamma+O(\gamma\epsilon).
\end{equation}
The probability of measuring parity one (i.e., a correct result) is
modified in two ways. First, the probability of reading  $111$ becomes
\begin{equation}
\mathbb{P}(p=111)=(1-\gamma)^{3}(1-\epsilon)^{3}=(1-\epsilon)^{3}-3\gamma+O(\gamma^{2},\gamma\epsilon).
\end{equation}
However, parity can also be one if the readout is $100$. This
can happen in two ways: 1) the first exists also in the case where
no decay has happened. This event occurs with a probability $(1-\gamma)^{3}p_{1}(1-\epsilon)\epsilon^{2}=p_{1}\epsilon^{2}+O(\epsilon^{2}\gamma)$.
2) The sequence $100$ can also occur due to decay without readout error.
\begin{equation}
\mathbb{P}(p=100)=(1-\gamma)\gamma(1-\epsilon)^{3}=\gamma+O(\gamma^{2},\gamma\epsilon).
\end{equation}
Clearly, $010$ and $001$ can also happen, but these sequences cannot happen
due to decay without readout error, so their role remains as it was
in the decay-free case. In total, the probability that the parity outcome
is \textquotesingle $1$\textquotesingle\ in a sequence of three measurements is
\begin{equation}
\mathbb{P}[par(3)=1]=(1-\epsilon)^{3}-3\gamma+\gamma+O(\gamma^{2},\gamma\epsilon)=(1-\epsilon)^{3}-2\gamma+O(\gamma^{2},\gamma\epsilon).
\end{equation}
In a sequence of five measurements, we have
\begin{equation}
\mathbb{P}(p=11111)=(1-\gamma)^{5}(1-\epsilon)^{5}=(1-\epsilon)^{5}-5\gamma+O(\gamma^{2},\gamma\epsilon).
\end{equation}
Now, the decayed physical qubit states that contribute to $par(5)=1$
are $ 11100$ and $10000$ and therefore
\begin{align}
\mathbb{P}(par(5)=1)&=\mathbb{P}(p=11111)+\mathbb{P}(p=11100)+\mathbb{P}(p=10000)\\
&=(1-\epsilon)^{5}-5\gamma+(1-\gamma)^{3}\gamma(1-\epsilon)^{5}+(1-\gamma)\gamma(1-\epsilon)^{5}\\
&=(1-\epsilon)^{5}-3\gamma+O(\gamma^{2},\gamma\epsilon).
\end{align}
More generally, for an amplification factor of $2j+1$
\begin{equation}
\mathbb{P}(par(2j+1)=1)=(1-\epsilon)^{2j+1}-(j+1)\gamma+O(\gamma^{2},\gamma\epsilon).\label{eq: par_gamma}
\end{equation}
When considering the mitigated values ($m>0$), we find
\begin{equation}
\sum_{j=0}^{m}a_{j}^{(m)}\mathbb{P}(par(2j+1)=1)=1+O(\epsilon^{m+1})-\frac{\gamma}{2}+O(\gamma^{2},\gamma\epsilon).\label{eq: mitigate decay no dmy}
\end{equation}
Interestingly, there is a decay effect, but it does not depend on
the order $m$. The longest measurement sequence in mitigation order
$m$ is $2m+1$ and, as expected, according to (\ref{eq: mitigate decay no dmy})
longer sequences are more susceptible to decay. The reason why the
stronger degradation does not appear in (\ref{eq: mitigate decay no dmy})
can be understood by examining the decay contributions to the mitigated
outcome ($m>0$)
\begin{equation}
-\gamma\sum_{j=0}^{m}a_{j}^{(m)}(j+1)=-\frac{\gamma}{2}\sum_{j=0}^{m}a_{j}^{(m)}(2j+1)-\gamma\frac{1}{2}\sum_{j=0}^{m}a_{j}^{(m)}=-\frac{\gamma}{2}.
\end{equation}
where the last transition was made by exploiting the Taylor coefficient
properties.

\subsection{Mitigation of the decay effect using dummy measurements}

The decay scales as $j+1$, where $2j+1$ is the number of measurements
in a sequence. Since this scaling can only be partially mitigated,
we suggest adding $j$ dummy measurements for every value of $j$
and using only the last $2j+1$ measurements to calculate the parity.
For example, to evaluate $par(3)$ four measurements
are carried out and the last three are used. In $par(5)$
the last five measurements out of seven are used, so on. Repeating
the previous calculation with $d(j)$ dummy measurements, we find
that
\begin{equation}
\mathbb{P}(par(2j+1)=1)=(1-\epsilon)^{2j+1}-(j+d(j)+1)\gamma.
\end{equation}
By selecting $d(j)=j$ the corrected decay scales as $(2j+1)\gamma$
which is a noise scaling that the Taylor coefficients mitigate to
zero. As a result, for $m>0$ we obtain
\begin{equation}
\mathbb{P}(1|q=1)^{parity+dummy\:meas.}=\sum_{j=0}^{n}a_{j}^{n}\mathbb{P}(par(2j+1)=1)=1+O(\epsilon^{m+1})+O(\gamma^{2},\gamma\epsilon).
\end{equation}
This technique is useful when the final  physical state of the qubit  is irrelevant
and when the spectator qubits are fully protected during the measurements.
However, since the physical state with dummy measurements decays as
$\Gamma,\Gamma^{4},\Gamma^{7},...$ where $\Gamma$ is the decay channel,
the Taylor coefficients will fail to properly mitigate the noise in
the space of the qubit as they are designed to mitigate $2j+1$  powers, i.e., $\Gamma,\Gamma^{3},\Gamma^{5}$. Meanwhile,
the other qubits that are not being measured also decay while being
idle. With dummy measurements, they undergo the same decay pattern
$\Gamma,\Gamma^{4},\Gamma^{7},...$ and will not be properly mitigated
as well.

\begin{figure}
\includegraphics[width=0.5\textwidth]{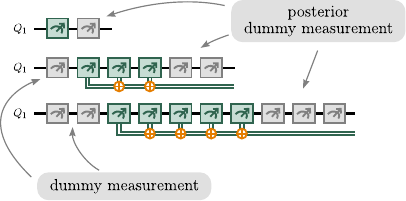}
\caption{To account for the decay and depolarization of the qubit during the
parity sequence, $j$ dummy measurements (left) are inserted before
the $2j+1$ measurement used for evaluating the $2j+1$ parity. While
this technique works well for terminating measurements, it leads to
a residual error in the state of the qubit. By adding $j+1$ posterior
dummy measurements (right), the scaling of this error becomes $2(2j+1)$
and therefore the mitigation successfully removes it in post-processing.
Spectator qubits must also undergo a $2(2j+1)$ delay so that their noise
will be mitigated as well. Due to the posterior measurements, the
effective decay rate to be mitigated in the qubit space increases
from $\gamma$ to $2\gamma$. This can be avoided by using the weighted
parity scheme. }\label{fig: dummy}
\end{figure}

In order to avoid this effect and employ dummy measurements in mid-circuit
measurements, we suggest adding $j+1$  "posterior dummy measurements" after the measurements used for evaluating $par(2j+1)$. As a result,
the decay noise scaling becomes $\Gamma^{2},\Gamma^{6},\Gamma^{10},...=(\Gamma^{2})^{1},(\Gamma^{2})^{3},(\Gamma^{2})^{5},...$
which fits the $2j+1$ noise scaling that Taylor mitigation properly
mitigates. This is illustrated in
Fig. \ref{fig: dummy}. The cost of the posterior dummy measurement protocol is the amplification
of the native noise from $\Gamma$ to $\Gamma^{2}$, or stated in
terms of rates $\gamma\to2\gamma$. This noise increase affects only
the qubit degradation noise. The measurement performance is exactly
the same as in the case without the posterior dummy measurement.

\subsection{Other types of noise on the state of the qubit}

According to the argument above, if we assume that the noise is mostly
local and consider only order $\gamma$, it suffices to focus
on single qubit noise. Since $\sigma_{y}$ is equivalent to $\sigma_{x}$
and $\sigma_{z}$ is equivalent to identity in terms of population
dynamics, there is only one extension to consider: a Pauli $\sigma_{x}$
noise, i.e., a bit-flip. This noise is important since, upon measurement
twirling, qubit decay noise becomes a bit-flip noise. We argue that
with order $\gamma$ the noise manifests in the same way as in
the decay case studied above. Therefore, it is also resolved by the
inclusion of dummy measurements as in the decay case.

To see this, consider the case where the decay rate is $\gamma_{\downarrow}$
and there is an additional excitation rate $\gamma_{\uparrow}$. For
bit flip errors it holds that $\gamma_{\downarrow}=\gamma_{\uparrow}$. Starting
initially in $q=1$, if a decay occurred after the first measurement the physical state of the qubit in three consecutive measurement is  $q=100$.
Unlike the decay-only case, after the first $\gamma_{\downarrow}$
event, there is a chance to go back to \textquotesingle $1$\textquotesingle \ so the final state can be $q=101$ or q=110. However, such events are
of order $\gamma_{\downarrow}\gamma_{\uparrow}$ and we neglect all
$O(\gamma^{2})$ events. In other words, for an initial state \textquotesingle $1$\textquotesingle \
and first-order analysis, bit-flip errors are indistinguishable from
amplitude damping errors. If the initial state is \textquotesingle $0$\textquotesingle, there will
be a major difference. In the decay case, there will be no dynamics
in the qubit space, since the qubit is already at the ground state.
However, when $\gamma_{\uparrow}\neq0$, there is a chance to make
a transition to the excited state. Yet, in $O(\gamma_{\downarrow},\gamma_{\uparrow})$,
this is indistinguishable from "decay to the excited state" and the
dummy measurement scheme will take care of this noise in the same
way. Thus, no modifications are needed to the dummy measurement scheme to address
 $\gamma_{\uparrow}\neq0$. It will equally work for
 decay noise, bit flip noise, or any combination thereof. 

\subsection{Mitigation of the decay effect using weighted parity}

The complete dummy measurement scheme, which includes posterior dummy measurements,
doubles both the execution time of the protocol and the decay noise
with respect to the basic parity protocol shown in Fig. \ref{fig: parity no decay}.
Although the decay noise is mitigated, this noise rate doubling increases the
sampling overhead. In this section, we introduce a more refined scheme
that overcomes this problem using a "weighted-parity" function. Although
the new scheme only involves a minor additional requirement on the
collected data, it is less intuitive and requires slightly more coding
effort than the dummy measurement scheme.

In Sec. \ref{subsec:The-impact-of decay}, we found that the reason
why the correction to the parity outcome is $(j+1)\gamma$ and not
$(2j+1)\gamma$ is the decay of the qubit into "left-aligned" states
with parity one e.g., $\{1000000,1110000,1111100\}$ for $j=3$
, that modifies the probability of classifying the state $q=1$ as
 \textquotesingle $1$\textquotesingle . Just to clarify, the state $1111111$ is considered as "centered"
and not left or right aligned. As an alternative to the dummy measurement
scheme, we suggest using a weighted parity function. This protocol
contains two steps. The first step is to calculate the $2j+1$ parity
in the usual way using $2j+1$ measurements, i.e., there are no dummy
measurements before or after. The parity outcome is used for the feedforward
and (or) for the readout. Operationally, the weighted parity protocol
is identical to the basic parity protocol shown in Fig. \ref{fig: parity no decay}, and the difference lies only in post-processing. Along with the result of the $2j+1$
parity, the measurement sequence $s_{i}$ of the $2j+1$ measurement
is recorded for each shot $i$. Thus, for each shot $i$ we obtain
a string $s_{i}$ of measurements sequence and an outcome $A_{i}$
for the observable of interest $A$. In feedforward scenarios, we have
\begin{equation}
A_{i}=\braOket A{K_{par(s_{i})}\ketbra{s_{i},q_{i}}{s_{i},q_{i}}K_{0}}{\rho_{0}}.
\end{equation}
where $K_0$ is the evolution operator in Liouville space \cite{gyamfi2020fundamentals,npjqiKIK,LayeredKIK} of the circuit that precedes the mid-circuit measurement and $ K_{par(s_{i})} $  is the evolution of the conditioned circuit that follows the measurement.
This quantity is evaluated for each noise amplification factor $2j+1$, so the collected data is  $s_{i}^{(2j+1)}$ and $A_{i}^{(2j+1)}$. In the
second step, instead of taking the standard average over different
shots in post-processing, we use the set $\{A_{i},s_{i}\}$ to calculate
the expectation value with amplified readout error:
\begin{equation}
\left\langle A\right\rangle _{W}^{(2j+1)}=\frac{1}{N_{shots}}\sum_{i}W(s_{i}^{(2j+1)})A_{i}^{(2j+1)},
\end{equation}
where the weight function $W$ is defined as follows
\begin{itemize}
\item For non-aligned states $W(s)=1$. This includes all ones and all zeros
states.
\item For right-aligned states $W(s)=2par(s)$.
\item For left-aligned states $W(s)=2(1-par(s))$. 
\end{itemize}
The $m$-th order mitigated expectation value is
\begin{equation}
\left\langle A\right\rangle _{mit}^{(m)}=\sum_{j=0}^{m}a_{j}^{(m)}\left\langle A\right\rangle _{W}^{(2j+1)}.
\end{equation}
Note that this choice of coefficients is not unique and further optimization
may lead to additional operational advantages. However, this choice
is simple and can be compactly described. To understand how this weighting
works, consider an amplification level of three, where the initial
qubit state is  initially in the \textquotesingle $1$\textquotesingle \ state.  The corresponding probabilities and weights
are shown in Table \ref{tab: weighted parity}. 
\begin{table}
\begin{tabular}{ccccc}
\toprule 
Readout sequence & Parity & Probability & $\text{Probability}+O(\gamma^2,\gamma\epsilon,\epsilon^{3})$ & W(s)\tabularnewline
\midrule
111 & 1 & $(1-\gamma_{\downarrow})^{3}(1-\epsilon)^{3}$ & $1-3\epsilon-3\gamma_{\downarrow}+3\epsilon^{2}$ & 1\tabularnewline
100 & 1 & $(1-\gamma_{\downarrow})^{3}(1-\epsilon)\epsilon^{2}+(1-\gamma_{\downarrow})\gamma_{\downarrow}(1-\epsilon)^{3}$ & $\epsilon^{2}+\gamma_{\downarrow}$ & 0\tabularnewline
010 & 1 & $(1-\gamma_{\downarrow})^{3}(1-\epsilon)\epsilon^{2}$ & $\epsilon^{2}$ & 1\tabularnewline
001 & 1 & $(1-\gamma_{\downarrow})^{3}(1-\epsilon)\epsilon^{2}$ & $\epsilon^{2}$ & 2\tabularnewline
000 & 0 & $(1-\gamma_{\downarrow})^{3}\epsilon^{3}+\gamma_{\downarrow}(1-\epsilon)^{3}$ & $+\gamma_{\downarrow}$ & 1\tabularnewline
011 & 0 & $(1-\gamma_{\downarrow})^{3}(1-\epsilon)^{2}\epsilon$ & $\epsilon-2\epsilon^{2}$ & 0\tabularnewline
101 & 0 & $(1-\gamma_{\downarrow})^{3}(1-\epsilon)^{2}\epsilon$ & $\epsilon-2\epsilon^{2}$ & 1\tabularnewline
110 & 0 & $(1-\gamma_{\downarrow})^{3}(1-\epsilon)^{2}\epsilon+(1-\gamma_{\downarrow})^{2}\gamma_{\downarrow}(1-\epsilon)^{3}$ & $\epsilon-2\epsilon^{2}+\gamma_{\downarrow}$ & 2\tabularnewline
\bottomrule
\end{tabular}

\caption{Given that the physical state of the qubit is initially \textquotesingle $1$\textquotesingle, the
left column shows the qubit state in three consecutive measurements.
Due to the decay rate $\gamma_{\downarrow}$, the qubit can decay to
\textquotesingle $0$\textquotesingle. The readout error rate is $\epsilon$ and processes of order
$O(\gamma^{2},\gamma\epsilon)$ are neglected where  $\gamma^2$ may refer to $\gamma_\downarrow^2$  or  $\gamma_\downarrow\gamma_\uparrow$  The standard parity
obtained by summing all the parity one probabilities with weight $W=1$
gives an error scaling in $\gamma$ that the Taylor mitigation coefficients
cannot mitigate since they are designed to mitigate the scaling of the
$\epsilon$ terms. When summing using the weights given in the right
column the correct scaling is obtained for both $\gamma$ and $\epsilon$, resulting in parity-based mitigation that is resilient to decay effects. Unlike the dummy
measurement this protocol is shorter and involves less noise before
mitigation.}\label{tab: weighted parity}

\end{table}

Note that $\gamma_{\uparrow}$ does not appear in the expressions
in Table \ref{tab: weighted parity} since we start with a state
\textquotesingle $1$\textquotesingle \ and consider only first-order decay processes. Similarly, when
starting with a state \textquotesingle $0$\textquotesingle, only $\gamma_{\uparrow}$ appears. The
standard parity is recovered by setting $W=1$ for all states. The
probability of measuring parity \textquotesingle $1$\textquotesingle \ (\textquotesingle $0$\textquotesingle) is the sum of the four top
(bottom) entries of the fourth column 
\begin{align}
\mathbb{P}(par(3)&=1)_{W=1}=1-3\epsilon-2\gamma_{\downarrow}+6\epsilon^{2},\\
\mathbb{P}(par(3)&=0)_{W=1}=3\epsilon+2\gamma_{\downarrow}-6\epsilon^{2}.
\end{align}
Assume that through feedforward, the parity leads to expectation value $A_{0}$
if $par(3)=0$, and to $A_{1}$ if $par(3)=1$. Therefore,
\begin{align}
\left\langle A\right\rangle _{W=1}^{(3)} & =A_{0}\mathbb{P}(par(3)=0)+A_{1}\mathbb{P}(par(3)=1)\nonumber \\
 & =A_{0}[\mathbb{P}(011)+\mathbb{P}(101)+\mathbb{P}(110)+\mathbb{P}(000)]\nonumber \\
 & +A_{1}[\mathbb{P}(111)+\mathbb{P}(100)+\mathbb{P}(010)+\mathbb{P}(001)]\nonumber \\
 & =A_{0}(3\epsilon+2\gamma_{\downarrow}-6\epsilon^{2})+A_{1}(1-3\epsilon-2\gamma_{\downarrow}+6\epsilon^{2})+O(\gamma^{2},\gamma_\downarrow\epsilon).\label{eq: A3W1}
\end{align}
Repeating the same calculation with $j\neq1$ amplification factors,
shows that the coefficients of $\gamma_{\downarrow}$ in $A_{0}$
and $A_{1}$ scales as $j+1$. As such, it cannot be mitigated using
Taylor mitigation.  So far we have restated the result of Eq. (\ref{eq: par_gamma}) in a feedforward setting. Next, we redo the calculation in Eq. (\ref{eq: A3W1})
with $W$ given by the right column of Table \ref{tab: weighted parity}
and obtain
\begin{align}
\left\langle A\right\rangle _{W=1}^{(3)} & =A_{0}\mathbb{P}(par(3)=0)+A_{1}\mathbb{P}(par(3)=1)\nonumber \\
 & =A_{0}[0\times\mathbb{P}(011)+1\times\mathbb{P}(101)+2\times\mathbb{P}(110)+1\times\mathbb{P}(000)]\nonumber \\
 & +A_{1}[1\times\mathbb{P}(111)+0\times\mathbb{P}(100)+1\times\mathbb{P}(010)+2\times\mathbb{P}(001)]\nonumber \\
 & =A_{0}(3\epsilon+3\gamma_\downarrow-6\epsilon^{2})+A_{1}(1-3\epsilon-3\gamma_\downarrow+6\epsilon^{2})+O(\gamma^{2},\gamma_\downarrow\epsilon).\label{eq: A3W}
\end{align}
Now the scaling of the $\gamma$ coefficient is $2j+1$ ($j=1$ in this example) and the Taylor
coefficients fully mitigate first order decay effects. This
should be repeated for the case where the initial physical state of
the measured qubit is \textquotesingle $0$\textquotesingle. However, the new table can be obtained
by setting $\gamma_{\downarrow}\to\gamma_{\uparrow}$ applying  a bit-wise NOT
transformation to the state, e.g., $100\to011$. All other
entries in the table remain the same and in the same position. The
weighted-parity mitigation works exactly as it did for the initial
state \textquotesingle $1$\textquotesingle. 

This method has the advantage that the scaling of the physical qubit noise scales as $2j+1$ and therefore it is automatically mitigated by applying
mitigation based on this scaling, e.g., the Taylor coefficients or the
adaptive coefficients in \cite{npjqiKIK}. Compared to the posterior
dummy measurement method, the native qubit error rate to be mitigated
is $\gamma$ and not $2\gamma$. As a result, weighted parity performs
better in mitigating qubit noise errors. The readout mitigation
performance is the same in both techniques. That being said, because of
its simplicity, we recommend the dummy measurements as a starting point
or for terminating measurements.

\subsection{Comparison to majority vote}

Having access to a sequence of measurements that measure the same
data, i.e., redundant noisy data, it is natural to ask how classical
error codes would perform. For a majority vote error correction in a sequence of length
$2m+1$ for a single qubit, we find that
\begin{equation}
\mathbb{P}_{maj\:vote}(1|q=1)=1+O(\epsilon^{m+1})-(m+1)\gamma.\label{eq: majority}
\end{equation}
In the absence of decay, the performance is equivalent to that of
Taylor coefficient mitigation, but majority vote incurs no sampling
overhead, which seemd as a key advantage over error mitigation method. However, Eq. (\ref{eq: majority}) shows that the decay strongly
degrades its performance. In majority vote the decay introduces a
bias of $(m+1)\gamma$ while in parity with dummy measurements the
correction is O($\gamma^{2})$. In the IBMQ experiments we have conducted,
we observed that even with three measurements ($m=1$) majority vote
leads to results that are worse than the raw unmitigated results, and
as $m$ increases, the performance linearly deteriorates as predicted by Eq. (\ref{eq: majority}). See Appendix
III for single-qubit experimental data.

\section{Reset-based readout error amplification and mitigation}\label{sec: Reset-based-readout-error}

\begin{figure}
\includegraphics[width=0.5\textwidth]{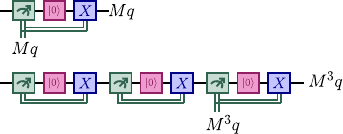}
\caption{Reset scheme: an alternative to the parity protocol for platforms
with high-fidelity reset, e.g., trapped ions. (top) A reset followed
by a conditional X gate ($\sigma_x$) ensures that the qubit state is $Mq$ even
if the measurement process completely destroys the state. (bottom)
By iterating this step, it is possible to achieve an $M^{{2j+1}}$ assignment matrix
as shown here for amplification factor three. Although the last reset
and X gate are not needed for the readout mitigation, in mid-circuit
measurement scenarios, they are crucial for mitigating the noise in
the qubits state. It also leads to the correct scaling of the noise
in the spectator qubits (not shown here).}\label{fig: Reset}
\end{figure}

The parity scheme works well when the readout error is pronounced
and the qubit degradation (e.g., decay) is mild. However, on some platforms,
such as trapped ions, the readout error is small, but the readout
process can substantially alter the qubit state. In these cases, we
suggest an alternative scheme that performs well when a high-quality
qubit reset operation is available, as in the case of trapped ions.
This scheme involves resetting each qubit to its measured value after
each measurement in the sequence. As shown in Fig. \ref{fig: Reset}, after each reset,
a feedforward operation is used for setting the qubit to the measured
state. This procedure effectively implements a measurement that imprints
the noisy measured outcome onto the qubit state. Consequently, the destructive
nature of each measurement plays no role. The outcome of applying
this procedure in a sequence of $2j+1$ measurements and using the last measurement as the readout outcome, is exactly equal to a readout with  a $M^{2j+1}$ assignment matrix.
Unlike the parity scheme, this relation holds for any $M$ not just
for twirled assignment matrices. In Sec. \ref{sec: Experimental-results} we present
ten-qubit experimental results from the Quantinuum H1 platform. It
is important to note that this scheme is not applicable in platforms
that use conditional reset where the reset is only as
reliable as the measurement itself (e.g., IBMQ). 

\section{Parity-based mitigation of state preparation errors}\label{sec: Parity-mitigation-prep-errors}

Preparation errors are sometimes assumed to be small and ignored without
proper justification. This can lead to errors in the final measured
quantities and inaccurate characterization of the assignment matrix. See \cite{haupt2025PrepMit} for a characterization of preparation errors using non-computational states.
Although our parity-based readout error mitigation does not involve characterization,
its outcome is not automatically resilient to preparation errors. 

Different platforms exploit different reset techniques. While trapped
ion computers typically exploit optical pumping or some refinements
of it, superconducting circuits typically use conditional reset. The
optical pumping fidelity is in general high and it appears that it
does not generate substantial preparation errors. Therefore, in this
work, we focus on mitigating preparation errors associated with conditional
reset. To the best of our knowledge, drift-resilient mitigation of
preparation errors has not been previously proposed or studied. 

\begin{figure}
\includegraphics[width=1\textwidth]{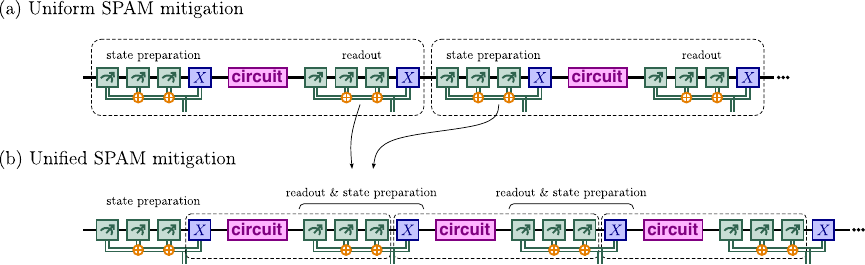}
\caption{(a) Treating the conditional reset element as a dynamic circuit, the readout error can be amplified using the weighted-parity scheme or the dummy measurement scheme (not shown). The combination of the original noisy circuit and (a) in post-processing yields a first-order error-mitigated initial state. (b) A more efficient realization is possible if parity-based mitigation is used for the terminating measurements. This "unified SPAM mitigation" mitigates both preparation and measurement error simultaneously without additional sampling overhead for the preparation errors. This is achieved by feed-forwarding the terminating outcome to a conditional X gate that executes a reset of the qubit for the next shot. In this scheme, the first circuit in each shot has to be slightly modified to account for the unknown initial state provided by the device. See main text for details.}\label{fig: prep}
\end{figure}

In a conditional reset, a qubit is measured, and a conditioned bit
flip is applied to the qubit. If the measured value is \textquotesingle $1$\textquotesingle, $\sigma_{x}$
is applied, and if it is \textquotesingle $0$\textquotesingle, no action is taken.
If the error in the X gate ($\sigma_{x}$) is negligible, the reset
error stems solely from the measurement error. By replacing the measurement
with $2j+1$ measurements and using the parity outcome for controlling
the X gate, we achieve a $2j+1$ amplification of the state preparation
error (Fig. \ref{fig: prep}a). When adding the results from $j=0,1,\ldots,m$
with the Taylor coefficients, the state preparation error is mitigated in
a drift-resilient manner. If the incoherent errors associated with
$X=R_{x}(\pi)$ are not negligible, then $R_{x}(\pi)$ should be replaced
by $R_{x}(\pi)[R_{x}(-\pi)R_{x}(\pi)]^{j}$ to amplify the incoherent
error by a factor of $2j+1$ as described in \cite{npjqiKIK}. To properly account for qubit decay effects, the weighted-parity or dummy measurements (with posterior measurement), should be used. In this approach, the default hardware-provider between "circuit shots" (dashed rectangular frames) is disabled and replaced by the parity-based state preparation. We point out that the term "shot" refers to one execution of a sequence of circuit shots. A batch of such executions is called a job.

In cases where the parity scheme is also used in the terminating measurements,
it is possible to combine the readout mitigation with the state preparation
mitigation and achieve a "unified SPAM" mitigation, as shown in
Fig. \ref{fig: prep}b (SPAM stands for state preparation and measurement errors). 
After implementing $2j+1$ measurements for the 
final readout in circuit shot number $k$, the parity outcomes activate control X gates that reset the state for the next circuit shot. This scheme is not only shorter than addressing the preparation
errors separately as in Fig. \ref{fig: prep}a, but it also has a reduced overall sampling overhead, since there is no measurement noise originating from measurements
dedicated to the reset of the initial state. 
Despite the noise advantage and the compactness of the unified-SPAM approach, 
it does involve an important subtlety. Inclusion of different amplification levels in the same job, for example, $n$ circuits of amplification $2j+1$ followed by $n$ circuit of amplification $2j+1$, should be avoided in the unified SPAM approach, since without proper treatment, it can lead to different transient dynamics in each amplification level that degrades the performance.  Since each job corresponds to different noise amplification, the level of drift resilience reduces to noise that changes on a time scale of several jobs (depending on the mitigation order) or slower.

\section{End-to-end mitigation and integration with other QEM methods}\label{sec: End-to-end-mitigation-and}

\subsection{End-to-end KIK-parity and KIK-reset mitigation}\label{subsec: KIK-parity}

The SPAM amplification by an odd factor that emerges in the parity and reset schemes is easily integrated
with the KIK noise amplification for gates \cite{npjqiKIK,LayeredKIK}.
A straightforward way to do so is to mitigate the SPAM errors for
each level of gate error amplification. An alternative method is to
amplify all errors by the same factor as shown in Fig. \ref{fig: parity no decay}b.
We refer to the latter scheme as "uniform amplification". The disadvantage
of the first scheme is that the total sampling cost is the product
of the gate mitigation sampling cost and the SPAM mitigation sampling
cost. Yet, this scheme facilitates the use of different mitigation
orders for the gates and for the SPAM errors. In the uniform amplification scheme, mild SPAM errors will add up to the gate noise. In general, it is recommended to use  the uniform amplification scheme unless there is a clear reason for not doing so.

By complementing the uniform amplification
scheme with twirling techniques for treating coherent errors, we obtain an end-to-end mitigation
strategy that addresses a wide variety of errors and is applicable
to any dynamic circuit (including error correction) in a drift-resilient
manner. In particular, we use pseudo twirling for non-Clifford multi-qubit gates, and randomized
compiling for two-qubit Clifford gates. The inclusion of dynamical decoupling for spectator qubits
is straightforward - the pulse inverse of the dynamical decoupling pulse sequence should be used
as any other gate \cite{LayeredKIK}. 

When using $2j+1$ error amplification, it is sometimes informally claimed that
it is unrealistic to use large amplification factors (beyond 5 or 7),
or that the duration of the circuits with the largest amplification
factor must remain below the coherence time of the device. However,
there is no such requirement or limitation. Even excessive noise amplification
holds additional valuable information on the manifestation of the
noise in the target circuit. Although the key approximation used in
\cite{npjqiKIK} that neglects the higher-order Magnus terms becomes
less justified at higher orders of mitigation, the layered KIK mitigation
\cite{LayeredKIK} solves this problem without incurring any additional
overhead. Therefore, the only limitation is the sampling overhead
(i.e., runtime) that is associated with higher-order mitigation. Yet,
since this end-to-end approach is drift-resilient and bias-free, shots can be executed
in parallel across different qubit sets, different processors or even
completely different hardware, and then combined in post-processing. 

\subsection{Integration with explicit matrix inversion methods}\label{subsec: Integration-inversion}

Assume that we have an assignment matrix $M$ that is already twirled.
In general $M$ is not in a tensor product form (local errors only), and it
also exhibits temporal drifts $M=M(t)$. We suggest exploiting an
approximate inverse $M_{approx}^{-1}$ using other methods to reduce sampling overhead and accelerate convergence of the parity method. The final result becomes drift resilient even if  $M_{approx}^{-1}$ is not drift resilient. $M_{approx}^{-1}$ can constructed using 
PEC \cite{hashim2025PECreadout,koh2024readout,Temme2024dynamicPEC} or using a tensor product of local matrix inversions $M_{loc}^{-1}=M_{1}^{-1}\otimes M_{2}^{-1}\otimes M_{3}^{-1}\ldots$
constructed, for example, from the product of local assignment matrices
$M_{i}$.
As a first step, we twirl $M_{approx}^{-1}$ so that it can be represented
in string form as 
\begin{equation}
M_{approx,s}^{-1}=\sum_{j_{i}..\in\{0,1\}}\beta_{j_{1},j_{2},j_{3}\ldots}\{j_{1},j_{2},j_{3}\ldots\}\oplus,
\end{equation}
where the 's' subscript means that we work again with string representation.
$M_{approx,s}^{-1}$ is defined to be time-independent so it
does not require relearning and updating. By applying the approximate
inverse to the noisy readout, the resulting effective assignment matrix
$M_{eff}(t)=M_{approx}^{-1}M(t)$ will be significantly closer to
the identity. However, $M_{eff}$ is still time-dependent due to drifts
and, in general, contains correlated noise. 

Since a product of twirled matrices is also twirled. We get that $2j+1$
parity is equivalent to $M_{eff}^{2j+1}$. Since $M$ is already
encapsulated in the noisy measurements $M_{approx}^{-1}$
is applied directly to the measured strings i.e., states in the measured
basis, generated by the noisy measurements. The application
of $M_{approx}^{-1}$ in a specific shot yields a quasi-probability.
This is inconvenient for mid-circuit measurements but can be resolved
by using the decomposition of $M_{approx,s}^{-1}$ into Pauli operations.
Fortunately, there is no need to do this for each measurement in the
sequence separately.  Due to the commutativity property of the XOR operation, multiplying each of the $2j+1$ measurement outcomes by $M_{approx,s}^{-1}$ and then calculating the parity is equivalent to first calculating the $2j+1$ parity and then multiply it by $(M_{approx}^{-1})^{2j+1}$.

The advantages of this hybrid approach are a reduction of the sampling overhead relative
to Taylor mitigation, on the one hand, and drift-resilience compared
to matrix inversion methods, on the other hand. If there is no twirling,
then applying $M_{approx}^{-1}$ in post-processing yields $(M_{approx}^{-1}){}^{2j+1}M^{2j+1}$
that is different from $(M_{approx}^{-1}M){}^{2j+1}$. Yet, according
to the analysis in \cite{LayeredKIK} for layered-based amplification,
the Taylor coefficient will mitigate this as well, and the uncorrected
orders for mitigation order $m$, $O(\epsilon^{m+1})$, will be smaller
compared to the case where $M_{approx}^{-1}$ is not used. For example,
in first-order mitigation if we write $M=e^{+\xi A},M_{approx,s}^{-1}=e^{-\xi A_{approx}}$
then the leading-order correction in the hybrid approach  using  $(M_{approx}^{-1}M){}^{2j+1}$ is $\frac{3}{2}\xi^{2}(A- A_{approx})^{2}$, and in the $(M_{approx}^{-1}){}^{2j+1}M^{2j+1}$ case it is $\frac{3}{2}\xi^{2}((A- A_{approx})^{2}-[A,A_{approx}])$. Since  
correlated errors are smaller than local errors, $[A,A_{approx}]$
is expected to be small, and the difference between the two protocols will be small as well. 

To conclude this section, we point out that in cases where native
readout error is highly asymmetric, or if it symmetric but the qubit
noise is asymmetric (e.g., decay noise), it is possible to combine
parity-based and reset-based readout mitigation with a selective flip
of the qubit and post-processing correction as studied in \cite{shirizly2025feedforwardMeasBias,Bauer2021BiasFlipMeas}.
This approach is not applicable after twirling that symmetrizes both
the readout noise and the qubit noise.

\section{Experimental results}
\label{sec: Experimental-results}

To experimentally test our findings, we have conducted various experiments
on the superconducting IBMQ platform and the trapped ions Quantinuum
platform. 
\begin{figure}
\includegraphics[width=1\textwidth]{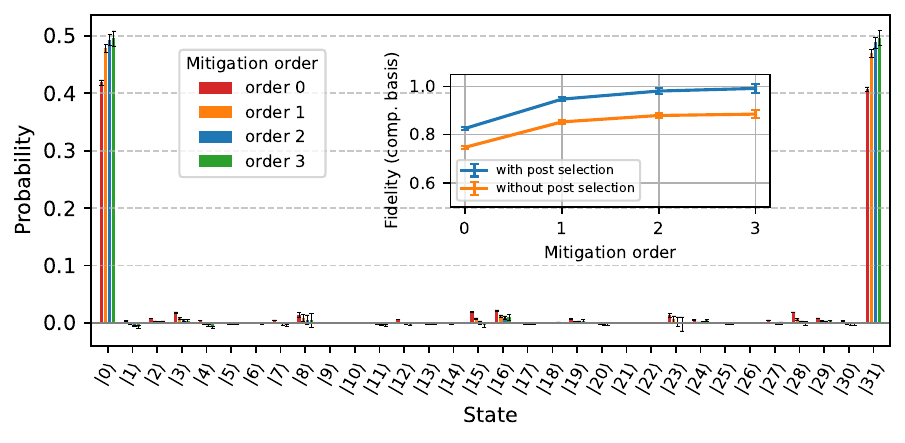}
\caption{End-to-end mitigation of a GHZ preparation using a dynamic circuit. Weighted parity is used for mitigating the mid-circuit measurement readouts and the terminating readout. The preparation error is mitigated by post-selection based on the three dedicated measurements. Randomized compiling/Pauli twirling is used for treating coherent errors in the CNOT gates and Layered-KIK is used for mitigating the incoherent error in the single and two-qubit gates. The blue curve in the inset shows the fidelity of the prepared state in the computational basis. To illustrate the impact of the preparation error we also show in the orange curve the fidelity if we do not use post selection to parity the initial state. It is possible to achieve faster convergence to fidelity one with the same experimental data using more advanced postprocessing that will be explored elsewhere.}\label{fig: GHZ exp}   

\end{figure}

\subsection{End-to-end mitigation of GHZ state preparation based on dynamic circuits}\label{subsec: GHZ-KIK-parity}

Our first experiment, show in Fig.  \ref{fig: GHZ exp}, implements a five-qubit dynamic circuit GHZ preparation as  presented in  \cite{baumer2024LongRange}. However, in our experiment, we apply the following: 1. Weighted parity for mitigating mid-circuit measurements and terminating measurements. 2. KIK method  for mitigating noise in single-qubit and two-qubit gates. 3. Pauli twirling for reducing coherent errors in CNOT gates and idling qubits.
Furthermore, instead of using the default hardware-provider reset to implement the mid-circuit reset, we have utilized the unified SPAM mitigation presented in Sec. \ref{sec: Parity-mitigation-prep-errors}. That is, the measurements that precede the reset operations are part of the reset process. Consequently, the reset operation is also mitigated.
To cope with the fiducial state ($00000$)  preparation errors, we use post-selection based on three additional measurements that take place before each circuit. Only shots that start with $000$ in each qubit are used in the post-processing mitigation. As a result, we discarded $27\%$ of the data. In Appendix II, we show that the three-measurement post-selection protocol reduces the fiducial state preparation error per qubit from $0.05$ to $10^{-5}$.

This experiment illustrates our "end-to-end" error mitigation approach that combines "Layered-KIK" and parity readout mitigation. The blue (orange) curve in the inset of Fig.  \ref{fig: GHZ exp}, shows the fidelity in the computational basis with (without) post-selection of the initial state.  Fidelity one is measured within the experimental uncertainty in third order mitigation.
While full quantification of GHZ state preparation requires additional measurements in different bases, our goal is restricted to demonstrating the improvement our mitigation approach provides in this scenario.   

We have used qubits 64,76, 75, 90, and  95 in the ibm\_strasbourg device. Each job sent to the backend contained forty circuit with different levels of noise amplification. Denoting these circuits by $K_{2j+1}$, the drift-resilient execution order  \cite{npjqiKIK,LayeredKIK} in each job is $K_1,K_3,K_5,K_7,K_1,K_3,K_5,\dots$ and each job contains 1000 shots. 

All measurements were twirled by random samples of multi-qubit Pauli
matrices. While $\sigma_{x}$ and $\sigma_{y}$ are indistinguishable
 in terms of population transformation, the inclusion of $\sigma_{y}$
and $\sigma_{z}$ reduces potential crosstalk effects. The same applies
for $\sigma_{0}$ and $\sigma_{z}$. In addition, $\sigma_{0}$ or
$\sigma_{z}$ were randomly inserted before each measurement to suppress
the potential buildup of coherent errors. Crucially, when executing
$\sigma_{z}$ or $\sigma_{y}$, it is important to implement the full
physical gate and not use virtual z ($z_{v}$) for $\sigma_{z}$ and
$\sigma_{x}z_{v}$ for $\sigma_{y}$. See \cite{npjqiKIK,Lidar2024virtualZ,2024PST}
for further details. A Misuse of virtual Z gates during twirling and
dynamical decoupling can result in an incomplete set of operations
that does not address all coherent errors.  The instances of  $K_{2j+1}$ in the same job undergo different twirling realizations. Since we have ten jobs, the total number of twirling realizations for each amplification level is 100. We note that considerably faster convergence can be achieved by applying new adaptive post-processing coefficients, which will be explored elsewhere.

The presence of noise drifts is easy to validate using the data collected during the experiment. In Appendix IV, we describe how the sequential measurements used in our method provide real-time diagnostic information on the quality of the qubits. The decay curves in this Appendix clearly show that noise in some of the qubits has changed substantially during the experiment.  Although, in principle, it is possible to use this information to  discard some of the data and get better results, our end-to-end mitigation protocol took care of all the noise drifts without discarding any data. Furthermore, in this experiment, dynamical decoupling was not used. Instead we have used sequential Pauli twirling cycles on the idling qubits. Dynamic decoupling can perform better than twirling and in larger experiments it is advisable to use it to reduce the noise load for error mitigation.

We have also tried to implement the "unified-SPAM" approach to handle the fiducial state preparation error  instead of the post-selection but we achieved only fidelity of $0.955\pm 0.004$. The reason could be the unknown classical processing time that may generate unaccounted delays. In addition, in the unified-SPAM execution we cannot change the amplification order from one circuit to another in each job, but only on a longer time scale. This leads to a suppressed drift resilience with respect to the circuit execution order used in the GHZ experiment with post-selection.

\subsection{Parity-based twenty-qubit experiment in superconducting circuits \label{sec: 20q fez}}

\begin{figure}
\includegraphics[width=8cm]{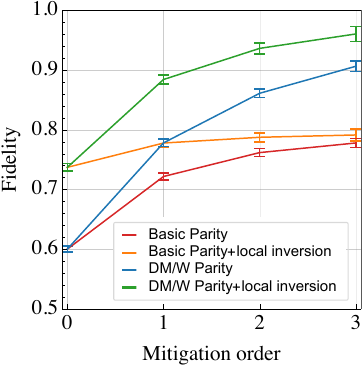}
\caption{A Twenty-qubit readout mitigation experiment carried out on the IBMQ
backend Fez. Measurements are twirled so the effect of the readout
noise on the fidelity is the same for all input states. Mitigation
order 0 corresponds to the raw data without mitigation. The basic
parity protocol described in Sec. \ref{sec: Parity-based-measurement}
is not resilient to decay effects and does not converge to the correct
result, as shown in the red curve. The blue curve shows the result
of the dummy measurement (DM parity) protocol or weighted parity (W-parity)
protocol. The DM and W parity results are undistinguishable in this
figure. The green curve demonstrates the clear advantage of the hybrid
protocol presented in Sec. \ref{subsec: Integration-inversion}. This
protocol combines time-independent local matrix inversion based on
the IBM website data on readout errors with dummy measurement (or
W) parity. Using an exponential fit and extrapolating to mitigation
order five we find that the mitigated fidelity is $0.977\pm0.027$.
The orange curve shows that the hybrid method does not work with the basic
parity amplification scheme that ignores qubit degradation effect
such as decay and bit-flips.}\label{fig: 20q exp}

\end{figure}

Figure \ref{fig: 20q exp} shows a twenty-qubit experiment conducted
 on the ibm\_fez backend. We executed a sequence of thirteen measurements
for each shot. The initial state is the all-zeros state. Since the
measurements are twirled, the readout fidelity should be state-independent.

As in the previous experiment, we used three measurements in the beginning of each circuit to mitigate the preparation error of the fiducial state through post-selection. The post-selection success rate (the percentage of shots that were not discarded) in the preset twenty-qubit experiment is 27.5\%.
For the weighted
parity, we used measurements numbered 4,4-6,4-9, and 4-10 to
evaluate par(1), par(3), par(5), and par(7), respectively. For the
dummy measurement protocol, 4,5-7,6-10 and 7-13 were used.

The regular parity scheme, which does not account for qubit decay,
performs poorly compared to the dummy measurement parity and the weighted
parity (blue curve). The dummy measurement parity and the weighted
parity are indistinguishable from each other at the scale of the plot.
By using the results of Sec. \ref{subsec: Integration-inversion}
and setting $M_{approx}^{-1}$ to be the local inverse (tensor product
inverse) constructed from the IBMQ website data we obtain the green
curve. Since the readout error is twirled, we also twirl $M_{approx}$
before the inversion. As evident from the figure, the convergence
is much faster when combining parity
mitigation with the local inversion. 

To demonstrate drift resilience, the data was collected at random times over a period of five days, and
$M_{approx}$ was evaluated based on the data available online on the final day.
Although $M$ changes over time during the five days, we have used the
same $M_{approx}$ for the entire dataset. This illustrates our claim
that $M_{approx}$ does not need to be updated.  Nonetheless, it is possible to have
a better starting point than simply using online data about measurement 
quality. When $M$ is twirled, $M_{approx}$ can be obtained from a
single experiment by recording the readout outcomes for the all-zeros
initial state. In particular, an uncorrelated tensor product version
of $M_{approx}$ can be easily calculated from the same data. Crucially,
the input state may contain residual preparation errors. Hence, it is
important to use post-selection or the methods described in Sec.  \ref{sec: Parity-mitigation-prep-errors}
for removing preparation errors. 

As a reference, the red curve shows the basic parity protocol that does not account for
qubit degradation. The orange curve shows the basic parity protocol with
local inversion. This demonstrates the importance of accounting for
the effects of qubit population degradation, such as decay and bit-flip
errors. 

By fitting an exponential curve to the "DM/W+local inversion"
fidelity data, we find that for the all-zeros input state, the extrapolated
fidelity is $0.975\pm0.027$ for order $m=5$. Testing the readout fidelity of other input states requires: 1) Post-selecting the all-zeros state and then applying a preparation circuit to create the new input state. 2) Applying KIK gate mitigation to the preparation circuit.  

In this experiment, we used two hundred twirling realizations, with
one thousand shots per realization. The qubits used in the experiment are qubits 0-19 of the ibm\_fez backend.

\subsection{Reset-based ten-qubit mitigation in trapped ions}

\begin{figure}
\includegraphics[width=0.5\textwidth]{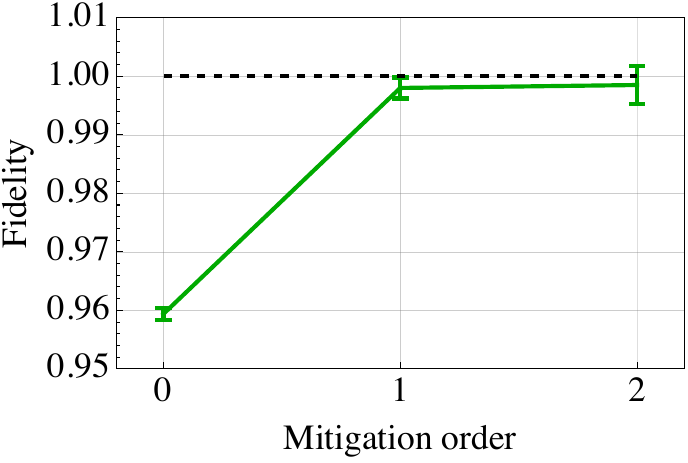}
\caption{Reset-based ten-qubit readout error mitigation experiment in the H1 trapped ion quantum computer of Quantinuum. The y-axis is the measured readout fidelity with
respect to the ideal all ones, $1111\ldots 1$  state, and the x-axis is
the mitigation order where order zero stands for the unmitigated results.
 }\label{fig: reset exp}
\end{figure}

In the next experiment, show in Fig. \ref{fig: reset exp}, we tested the reset scheme described in Sec
\ref{sec: Reset-based-readout-error} in the Quantinuum H1 trapped
ion quantum computer. The experiment consisted of ten qubits, each
of which was measured and reset nine times sequentially. No twirling was
used in this experiment. The results show a removal of the noise bias
already in first order mitigation.
We have also executed a variant in which two measurements were used to post-select the initial state and quantify the preparation error via the post-selection rate. We observed that post-selecting only shots that start with $00$ has little effect on the outcome. This is consistent with our assumption that this device performs a high-quality reset.

We point out that achieving convergence in first-order mitigation, as shown in Fig. \ref{fig: Reset}, requires utilizing the adaptive post-processing coefficients described in \cite{npjqiKIK}. A more general method will be studied elsewhere.
Unlike the parity method, the reset method is based on real-time feedforward, and therefore
this experiment demonstrates real-time capabilities of the H1 backend. Finally, we point out that when experiments with a higher number of qubits on this platform are considered, it is likely that more detailed information on the maximal simultaneous measurements and the shuttling used during the readout of a large number of qubits will become important. 

\section{Conclusion}

Resilience to time drifts is a highly desired feature of error mitigation
protocols, as it enables bias-free results in real devices where the
noise parameters change during the course of the experiment. Drift resilience enables the collection of a larger number of shots without introducing systematic bias, thereby reducing statistical error bars and, in turn, facilitating stronger error mitigation than was previously possible. 
To date,
Drift resilience has been limited to specific mitigation protocols for gate errors.
By unraveling a simple yet powerful relation between parity and readout
noise amplification, we have shown how to achieve drift-resilience
in the mitigation of measurement and preparation errors. In mid-circuit
measurements, our protocol also mitigates the degradation of the qubit
state during the measurement process, including that of spectator qubits. We
 have also shown, both analytically and experimentally, that the degradation
of the qubit state during the measurement prevents the use of simpler
classical error correction schemes, such as majority voting.  It is worth noting that the underlying principles in the weighted-parity (or dummy measurements) mitigation are essentially classical, and as such, this approach may find uses in classical systems as well. 

In addition to integration with QEC and, more broadly, with dynamic
circuits, the tools presented in this work can be used for other tasks
as well. One example is process tomography, which is crucial for characterizing
gate performance and quantifying the impact of various error mechanisms, such as crosstalk, decoherence, coherent errors, and more. However,
in addition to being time-consuming, process tomography is also challenging
due to SPAM errors. Gate-set tomography is a method that enables the
extraction of SPAM-free gate parameters, as well as the SPAM parameters.
However, gate-set tomography is even more time-consuming than
 process tomography. By employing parity-based or reset-based SPAM mitigation,
it is possible to characterize gates using standard process tomography
instead of gate-set tomography. If preparation errors are of interest,
parity-based measurement mitigation can be applied to the fiducial state
to measure the preparation errors directly. 

Another example of an application, described in Appendix
IV, is real-time diagnostics of qubit quality. Qubit quality
can degrade significantly during the experiment. In superconducting
circuits, this can occur due to the emergence of two-level system
defects \cite{kim2023evidence,kim2024IBMstabilizeNoisePEC}, and
in trapped ions or Rydberg atoms, it may result from ion or atom loss.
As shown in Appendix IV, by monitoring the average populations in each
measurement in the parity (or reset) sequence, it is possible to identify
faulty qubits. In this work, we have employed this approach to verify the existence of noise drifts. 

Notably, in the present work, the parity and reset measurement protocols were not optimized in any way. For example, a short delay between the measurements in the parity protocol might improve the results by removing residual oscillator ring-down effects. Furthermore, the duration of each measurement was optimized by the hardware providers for a readout based on a single measurement but other choices might be preferable when using parity or reset-based mitigation. An additional improvement in the parity scheme might be obtained by classifying ambiguous results as "unknown" instead of \textquotesingle $0$\textquotesingle \ or \textquotesingle $1$\textquotesingle. As a result, the probability of reading \textquotesingle $0$\textquotesingle \ or \textquotesingle $1$\textquotesingle \ will be less than one, i.e., leakage error. The parity scheme can mitigate leakage error, and preliminary tests suggest that this option is better than misclassifying and then correcting it with parity mitigation. The results may be further improved by classifying level \textquotesingle$2$\textquotesingle\ of the transmon and applying a pulse that brings it back to the computation space. The parity algorithm will experience this process as a decay event and mitigate it.

As we have demonstrated, the parity schemes, either weighted parity or dummy measurements,
and the reset scheme, can be combined with other methods: 1) other
readout mitigation methods can be used to reduce the sampling cost,
while the parity and reset schemes ensure drift-resilience. 2) They
can be integrated with gate error mitigation. In particular, integration
with the Layered KIK method is most natural, as it is also drift-resilient
and based on gate error amplification with the same amplification
factors used in the parity and reset methods. It is interesting
to experimentally explore the extent to which the end-to-end mitigation
approach (SPAM+gate errors) for dynamic circuits improves performance 
in emerging quantum error correction experiments. A successful demonstration
will pave the way for a symbiotic relationship between QEM and QEC,
that outperforms each method individually.  
\begin{acknowledgments}
We thank Yotam Shapira for clarifications on
the trapped ion readout schemes, and Liran Shirizly for suggesting
the use of post-selection for the initial state. RU is grateful for support from the Israel Science Foundation (Grants
No. 2556/20 and 2724/24). The support of the Israel Innovation Authority
is greatly appreciated. We acknowledge the use of IBM Quantum Credits for this work. The views expressed are
those of the authors, and do not reflect the official policy or position of IBM or the IBM Quantum team. We thank Qarakal Quantum for their support in conducting the experiments.
\end{acknowledgments}

\section*{Appendix I - why parity works even without measurement twirling}

Let $v_{i}$ be a computational basis vector in the measurement basis in an $n$ qubit system.
Without loss of generality, we consider the system to be in a state
$q=v_{1}$. Since $v_{i}^{T}v_{j}=\delta_{ij}$ a general confusion matrix can be written as
\begin{equation}
M=\sum_{i=1}^{2^{n}}p_{i,j}v_{i}v_{j}^{T}.
\end{equation}
Given that the actual state of the qubit is $v_1$ the average readout result will be
\begin{equation}
\left(\sum_{i=1}^{2^{n}}p_{i}v_{i}v_{1}^{T}\right)v_{1},
\end{equation}
where $p_{i}=p_{i,1}$ is the probability of measuring $v_{i}$ given that the qubits are in state $v_1$. When the noise
is weak, $p_{1}$ is close to one and $p_{i\neq1}\ll1$. We can list
the different events, their probabilities and parities. For three
consecutive measurements, we have

\begin{align*}
 &  &multiplicity &  &parity & & probability\:\:\:\:\:\\
\hline \{v_{1},v_{1},v_{1}\} &  & \left(\begin{array}{c}
3\\
0
\end{array}\right) &  & par(v_{1}) &  & p_{1}^{3}\\
perm\{v_{1},v_{1},v_{i\neq1}\} &  & \left(\begin{array}{c}
3\\
1
\end{array}\right) &  & par(v_{i},v_{1},v_{1})=par(v_{i}) &  & p_{i}p_{1}^{2}\\
perm\{v_{1},v_{i\neq1}v_{j\neq1}\} &  & \left(\begin{array}{c}
3\\
2
\end{array}\right) &  & par(v_{1},v_{i},v_{j}) &  & p_{i}p_{j}p_{1}\\
perm\{v_{l\neq1},v_{i\neq1}v_{j\neq1}\} &  & \left(\begin{array}{c}
3\\
3
\end{array}\right) &  & par(v_{l},v_{i},v_{j}) &  & p_{i}p_{j}p_{l}
\end{align*}
where by $ par(v_{l},v_{i},v_{j})$ we mean bit-wise par(3) for each qubit. For five consecutive measurements

\begin{align*}
 &  & multiplicity &  & parity &  & probability\:\:\:\:\:\\
\hline \{v_{1},v_{1},v_{1},v_{1},v_{1}\} &  & \left(\begin{array}{c}
5\\
0
\end{array}\right) &  & par(v_{1})=v_{1} &  & p_{1}^{5}\\
perm\{v_{1},v_{1},v_{1},v_{1},v_{i\neq1}\} &  & \left(\begin{array}{c}
5\\
1
\end{array}\right) &  & par(v_{i},v_{1},v_{1},v_{1},v_{1})=par(v_{i})=v_{i} &  & (p_{i})p_{1}^{4}\\
perm\{v_{1},v_{1},v_{1},v_{i\neq1},v_{j\neq1}\} &  & \left(\begin{array}{c}
5\\
2
\end{array}\right) &  & par(v_{1},v_{1},v_{1},v_{i},v_{j})=par(v_{1},v_{i},v_{j}) &  & (p_{i}p_{j})p_{1}^{3}\\
perm\{v_{1},v_{1},v_{l\neq1},v_{i\neq1},v_{j\neq1}\} &  & \left(\begin{array}{c}
5\\
3
\end{array}\right) &  & par(v_{i},v_{j},v_{l},v_{1},v_{1})=par(v_{i},v_{j},v_{l})  &  & (p_{i}p_{j}p_{l})p_{1}^{2}\\
\vdots &  & \vdots &  & \vdots &  & \vdots
\end{align*}
The correct outcome $par(v_{1})$ has probabilities that scales as $p_{1}^{1},p_{1}^{3},p_{1}^{5},\ldots$ as the noise amplification increases. Thus, from the  Taylor coefficients properties \cite{LayeredKIK}, in each amplification factor the mitigated probability is
\begin{equation}
\mathbb{P}(v_{1})=\sum_{j=0}^{m}a_{j}^{(m)}p_{1}^{2j+1}=\sum_{j=0}^{m}a_{j}^{(m)}e^{(2j+1)\epsilon}=1+O(\epsilon^{m+1}),
\end{equation}
where $\epsilon=\ln p_1 $. We can also study the mitigation effect on one error events
\begin{align}
\mathbb{P}(\text{one error}) & =\sum_{j=0}^{m}a_{j}^{(m)}\left(\begin{array}{c}
2j+1\\
1
\end{array}\right)p_{i}p_{1}^{2j}=\frac{p_{i}}{p_{1}}\sum_{j=0}^{m}a_{j}^{(m)}(2j+1)p_{1}^{2j+1}=\frac{p_{i}}{p_{1}}\frac{d}{dp_{1}}\sum_{j=0}^{m}a_{j}^{(m)}p_{1}^{2j+1}\nonumber \\
 & =\frac{p_{i}}{p_{1}}\frac{d}{dp_{1}}(1+O(\epsilon{}^{m+1}))=O(p_{i}\epsilon{}^{m}),
\end{align}
where we have used the Taylor coefficients properties \cite{LayeredKIK}. Using higher-order derivatives, we find
\begin{align}
\mathbb{P}(\text{two errors}) & =O(p_{i}p_{j}\epsilon{}^{m-1}),\\
\mathbb{P}(\text{three errors}) & =O(p_{i}p_{j}p_{l}\epsilon{}^{m-2}).
\end{align}
In the main text, we showed that for twirled measurements, $2j+1$
parity is equivalent to $M^{2k+1}$. For a general non-twirled $M$,
the two operations are different. However, the derivation above shows
that the error mitigation still works correctly when using
parity since what matters is the scaling of the noise, and not its exact
form. 

\section*{Appendix II - Post-selection mitigation of preparation errors}

Let the probability of initially being in state \textquotesingle $1$\textquotesingle \ due to a preparation
error be denoted by $x$. For simplicity, we neglect the decay since
it only improves the result. Using Bayes law
\begin{equation}
\mathbb{P}(q=1|p=000)=\frac{\mathbb{P}(p=000|q=1)\mathbb{P}(q=1)}{\mathbb{P}(p=000)}=\frac{\epsilon_{1\to0}^{3}x}{(1-x)(1-\epsilon_{0\to1})^{3}+x\epsilon_{1\to0}^{3}}\simeq\frac{x\epsilon_{1\to0}^{3}}{(1-x)}.
\end{equation}
For $x=\epsilon_{10}=0.05$ we find that $\mathbb{P}(q=1|p=000)=8\times10^{-6}$
that is much smaller than the experimental accuracy. As a result,
$\mathbb{P}(q=0|p=000)=1-\mathbb{P}(q=1|p=000)$ is very close
to one, and a reliable \textquotesingle $0$\textquotesingle\ \ input state is achieved. With two measurements
for post-selection we get $\mathbb{P}(q=1|p=00)\simeq1.5\times10^{-4}.$
The sampling overhead is $1/\mathbb{P}(p=000)\simeq1-x-3\epsilon_{0\to1}$.

\section*{Appendix III additional experiments}

\begin{figure}
\includegraphics[width=0.5\textwidth]{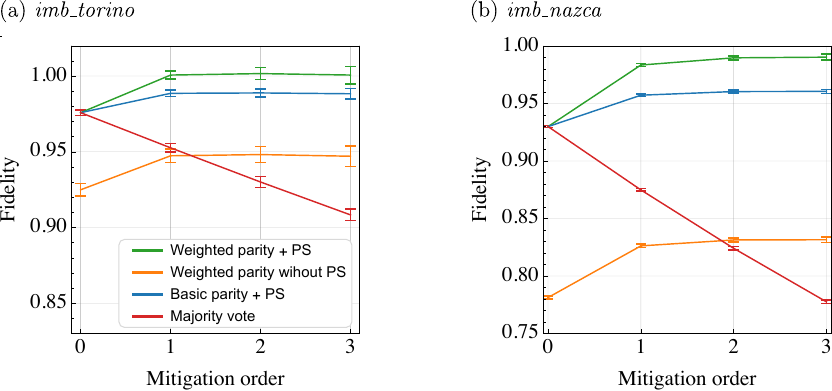}
\caption{Single qubit experiments on the IBMQ backends Torino (a) and Nazca (b). In all the backends we checked (including Sherbrooke and Brisbane)
it was crucial to use post-selection of the initial state to reduce
the native preparation errors. The green and orange curves show the
results of using weighted-parity with and without post-selection respectively.
The blue curves show the result of using the basic parity protocol
that does not account for qubit degradation. Although the manifestation
of noise is different in different backends, all the backends we checked
show strong and consistent mitigation when using both post-selection
and weighted parity (or dummy measurements). As a comparison, we also
plot in the red curve the majority vote outcome that can be applied
to the same data. However, due to the pronounced decay it produces results
that are worse than the unmitigated results. The linear degradation
is expected from Eq. (\ref{eq: majority}). Qubit no. 92 (63) was
used in ibm\_torino (ibm\_nazca).}\label{fig: 1q exp}
\end{figure}

In our first experiments, shown in Fig. \ref{fig: 1q exp}, we performed single-qubit readout mitigation across
various IBMQ backends. The preparation error treatment is based on
post-selection as in the main text experiment. There is one difference
though. In the early single qubit experiments, to create a more reliable \textquotesingle $1$\textquotesingle \ state
initial condition, we first applied an X gate and then post-selected
$111$ instances. Due to the metastable nature of the \textquotesingle $1$\textquotesingle \ state, this
methodology is not as effective as first post-selecting $000$, and
then applying the X gate to create the desired initial state. Furthermore,
no measurement twirling or forced decoherence were used in these early
single-qubit experiments.

The readout noise for a single qubit is sufficiently weak such that only
a single mitigation order was needed, and the second-order results
show stability and consistency with the first order.  In these examples, dummy measurements were used, but the results with weighted parity are nearly identical.

As discussed in the main text, in the absence of decay, a majority vote
yields similar performance to those obtained by our parity approach.
However, in agreement with Eq. (\ref{eq: majority}), we observe that due to decay effects, majority vote exhibits linear
degradation with respect to the unmitigated results (red curve in Fig. \ref{fig: 1q exp}).

\section*{Appendix IV - Real-time qubit diagnostics}

The information obtained from the consecutive measurements of the parity and reset protocols can also
be exploited to monitor the proper operation of the qubits. During our experiments,
we observed cases where some qubits behaved in a rather peculiar manner for
a certain time and then behaved normally again. Such behavior can
arise in superconducting qubits due to two-level system defects \cite{kim2023evidence,kim2024IBMstabilizeNoisePEC}.
A simple indication of the qubit's well-being is the evolution of
its population in the sequence of measurements used for the parity
evaluation. Consider a qubit that starts in state 1. If the noise channel
on the qubit (e.g., decay) is $K$, we expect that the average population
after j measurements will be $\braOket 1{MK^{j}}0$, so the population
dynamics as a function of $j$  should be a mixture of exponentials.  

While the input state is, in general, not known, it is possible to
post-select the states that start with \textquotesingle $1$\textquotesingle \ in a given qubit and monitor
the decay of this subset of states. For example, if the measurement
is twirled so that it preserves the fully mixed state, and the input
state is a fully mixed state, then no decay will be observed. However,
by post-selecting the \textquotesingle $1$\textquotesingle\ \ state, the decay to a fully mixed state can be observed. There is no need
to post-select the all-ones state for all qubits. The decay curves can be generated for each
qubit separately by tracing out the other qubits.

\begin{figure}
\includegraphics[width=0.5\textwidth]{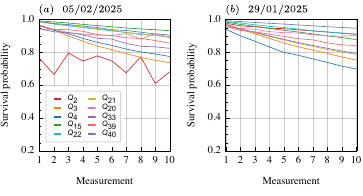}
\caption{Real-times diagnostics of qubits in ibm\_brisbane during the execution of a target
circuit. The parity scheme measurements are executed for each qubit
either for mid-circuit measurements and for terminating measurements.
The curves show the average population as a function of time. In this
example, all the qubits are initially in the \textquotesingle $1$\textquotesingle \ state. In the absence
of qubit decay, the curve should be more or less flat even  when
the measurement error is pronounced. The slope can be used to estimate
the $T_{1}$ amplitude damping noise that takes place during each
measurement. Apart from detecting changes in the decay rates this
method can also be used to detect qubits that are out of order as
shown by the bottom curve. While the initial state may not be \textquotesingle $1$\textquotesingle \
for a given qubit it is possible to post-select the shots in which
this is the case (see text). }\label{fig: daignostics}
\end{figure}

Figure \ref{fig: daignostics} depicts the decay curves of the same
qubits on the ibm\_brisbane device at two different times. In Fig. (a)
, qubit no. 2 is clearly behaving in an unexpected manner. However, a few
days earlier, this qubit behaved like all the other qubits, as shown in (b).
Such information can be used either in real-time to flag problems during
 execution or to identify time segments in post-processing where 
the qubits do not perform as expected, allowing for the discarding of data from these
 segments. We conclude that the consecutive measurements in the
parity scheme provide a valuable diagnostic tool without any additional
overhead. Unlike many other benchmarking tools, this tool directly  applies
to the data from the target experiment.

\begin{figure}[H]
\includegraphics[width=1\textwidth]{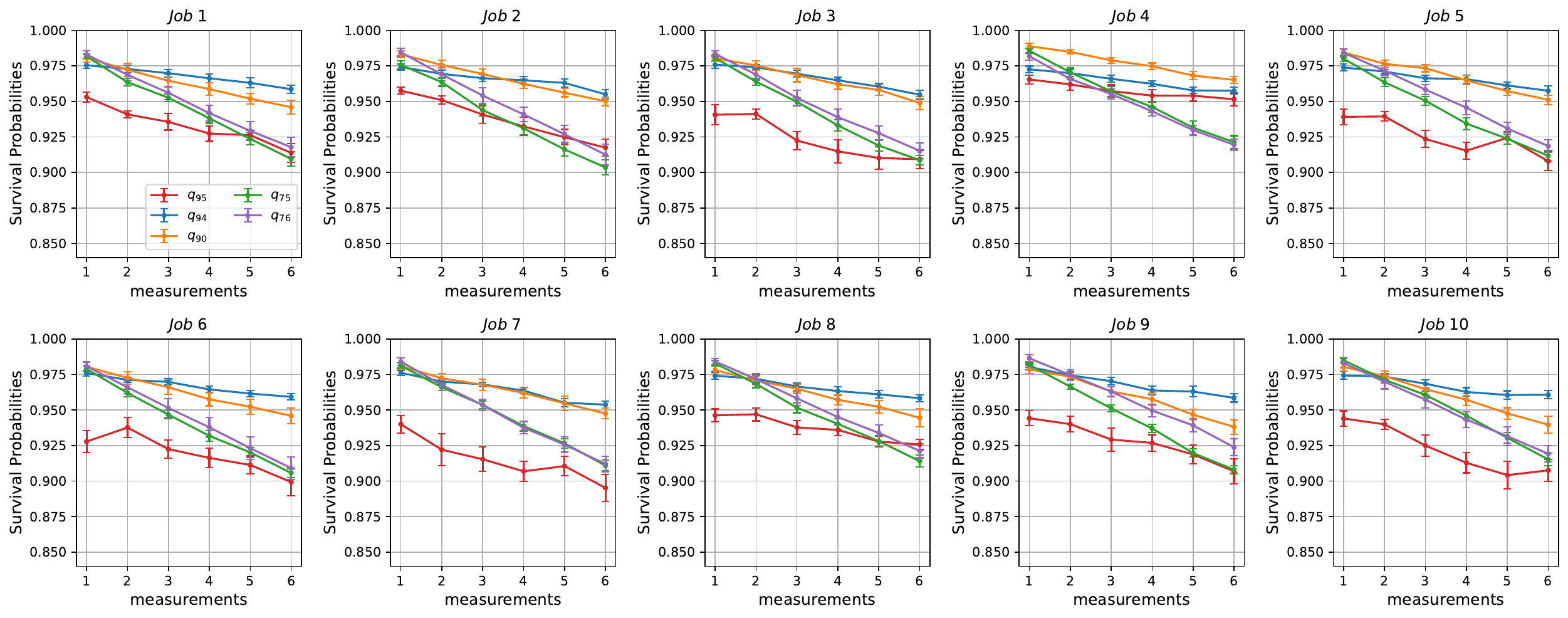}
\caption{The data in the GHZ experiment in the main text was collected in ten sequential jobs. The subplots show the decay curves in different jobs. The changes in the decay curves over time are a clear indication of temporal noise drift on a time scale of a single job (thousand shots of forty circuits, about 16-103 seconds with queuing and processing time, quantum runtime is 13 sec). Since these curves are generated from the same data used for the parity readout mitigation, this diagnostic tool does not incur any time overhead.}
\label{fig: diagnos GHZ}
\end{figure}

As a second example, in Fig. \ref{fig: diagnos GHZ} we show the decay curve of the ten sequential jobs used in the GHZ experiment presented in Sec. \ref{subsec: GHZ-KIK-parity}. 
The change in the decay curves from job to job demonstrates the presence of noise drifts. In particular, qubit no. 95 (red curve) exhibits a stronger noise drift compared to the other qubits. 

\begin{figure}[H]
\includegraphics[width=1\textwidth]{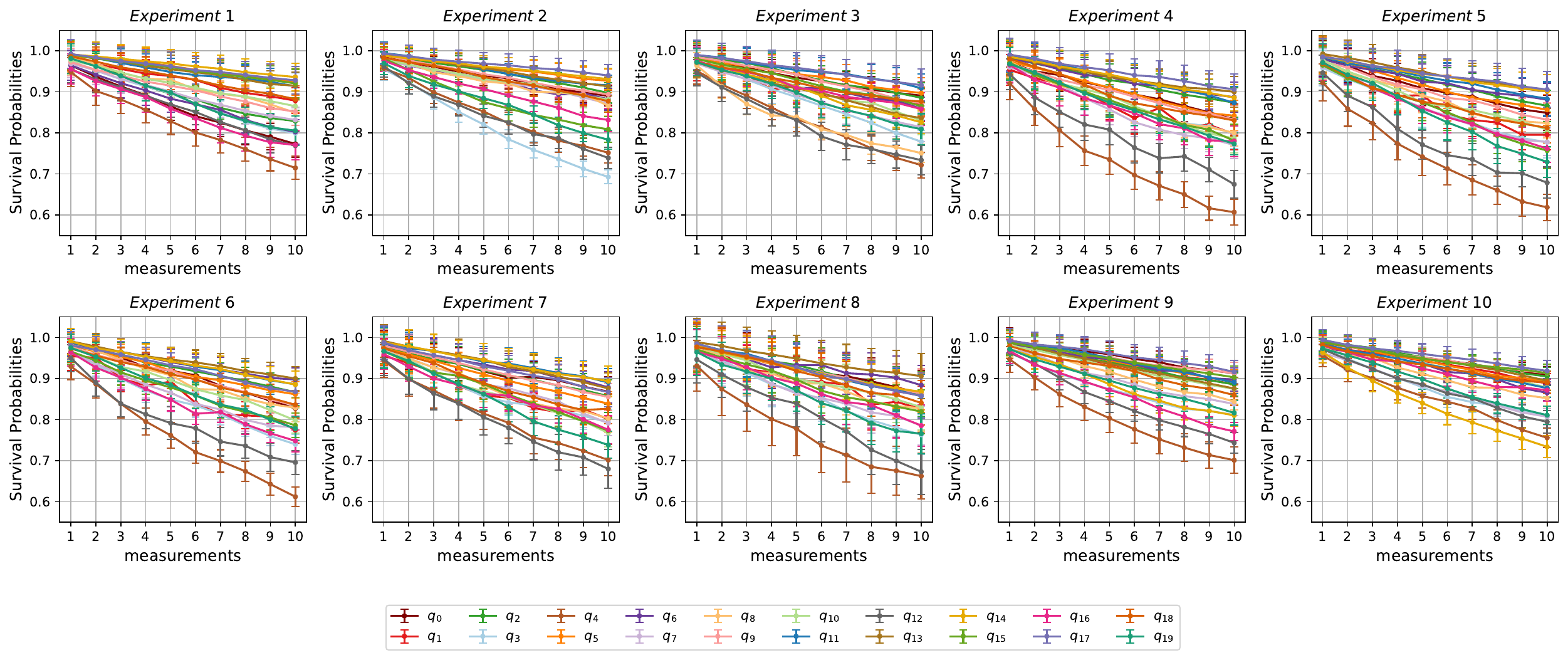}
\caption{The data in the twenty-qubit experiment in the main text was collected in ten non-sequential experiments over five days. Each experiment contained ten sequential jobs. Each subplot correspond to a different experiment. 
\label{fig: 20q decay}}
\end{figure}

Finally, in Fig. \ref{fig: 20q decay} we present the decay curves for the 20-qubit experiment in the main text \ref{sec: 20q fez}  where noise-drifts can also be observed. The experiments were carried out in random times over five days.

\bibliographystyle{apsrev4-1}
\bibliography{Refs_Parity}

\end{document}